\documentstyle[aps,prb,preprint]{revtex}
\begin{document}
\draft
\title{A Theory of the Longitudinal and Hall Conductivities of the
Cuprate Superconductors}
\author{Branko P.\ Stojkovi\'c and David Pines}
\address{Department of Physics and Materials Research
Laboratory,\\
1110 West Green Street, University of Illinois, Urbana, IL 61801
}
\date{\today}
\maketitle
\begin{abstract}
We establish the applicability to transport phenomena in the cuprate
superconductors of a nearly antiferromagnetic 
Fermi liquid (NAFL) description of
the magnetic interaction between planar 
quasiparticles by using it to obtain the
doping and temperature dependent resistivity and Hall conductivity seen
experimentally in the normal state.  Following a perturbative
calculation of the anisotropic (as one goes around the Fermi surface)
quasiparticle lifetimes which are the hallmark of a NAFL, we obtain simple
approximate expressions for the longitudinal, $\sigma_{xx}$, and Hall,
$\sigma_{xy}$, conductivities which reflect the magnetic crossovers seen
experimentally as one varies the doping level and temperature.  We present a
simple phenomenological model for the variation in mean free path around the
Fermi surface, and use this to extract from experiments on $\sigma_{xx}$ and
$\sigma_{xy}$ quasiparticle lifetimes in the hot (strongly coupled
quasiparticle) and cold (weakly coupled quasiparticle) regions of the Fermi
surface which are consistent with the perturbation theory estimates.  We improve
upon the latter by carrying out direct numerical (non-variational) solutions
of the Boltzmann equation for representative members of the 
YBa$_2$Cu$_3$O$_{6+x}$ and La$_{2-x}$Sr$_x$CuO$_4$ systems,
with results for transport properties in quantitative agreement with
experiment. Using the same numerical approach we study the influence 
of CuO chains on the a-b plane anisotropy and find results in agreement 
with experimental findings in YBa$_2$Cu$_4$O$_8$. 
\end{abstract}

\narrowtext
\vfill\eject
\section{Introduction}
\label{sec:intro}
\typeout{into on page \thepage}

The peculiar normal state properties of High
Temperature Superconductors (HTS) have
been extensively studied over the past decade in the expectation
that
their understanding
will likely reveal the mechanism of superconductivity. Yet no
consensus
has
emerged as to their physical origin.
Indeed, it is striking that
so much work has been done on understanding a variety of
complicated
experimental situations, while rather basic measurements, such
as the
d.c. transport, are not well understood.
So far it has been clearly established\cite{pines95}
and well documented in the
literature\cite{charlie,ong,iye}
that practically all normal
state properties in these  materials are anomalous with respect
to
the behavior of
standard Landau Fermi Liquids (FLs). This, in turn, poses the
question of whether Landau FL theory is applicable to
HTS.\cite{pwa} A variety of proposals based on
non-Fermi Liquid\cite{chakra,lee,altshuler} (NFL) or marginal
FL\cite{varma}
theories have emerged which have had
considerable success in explaining a number
of specific
experimental results. However, at present
no single theory has been able to
account for {\em all} the anomalies found in the normal state
properties
of the cuprates.

In this paper we focus on  the planar resistivity and the
Hall effect, 
which have drawn particular attention in the HTS community,
due to their arguably ``contradictory'' behavior:\cite{pwa-hall}
The
temperature dependences of the Hall angle, $\theta_H$, and the
longitudinal resistivity, $\rho_{xx}$, are so vastly different
that
the Hall coefficient, $R_H$,
in practically all cuprates is a strong function of
temperature, especially near optimal doping. This phenomenon by
itself,
is rarely observed in ordinary FLs.\cite{hurd}
It is also found that all HTS
near optimal doping have $\rho_{xx}$
proportional to the temperature, $T$, in a wide region of
interest.
Hence it appears that the transport properties in the presence
of
magnetic field as well as in its absence require
two separate
scattering mechanisms, and presumably two separate relaxation
rates.
In fact this is the only common scheme among the competing
theories
for the explanation of the anomalous transport in HTS materials.

We begin with a brief review of the theoretical proposals.
Anderson, who first drew attention to the problem of explaining
the unusual magnetotransport in the normal state,\cite{pwa-hall}
has proposed a scenario based on a NFL ground state. In his
scenario
the two rates observed in experiment are attributed to
scattering of two distinct many-body excitations, {\em spinons}
and {\em holons}. The existence of these excitations is well
established in 1D Luttinger liquids and Anderson has put forth
a non-trivial generalization of this concept to the quasi-two
dimensional
system found in cuprate superconductors.
Strictly speaking one can talk about a NFL in any
system at, or near half-filling,
which has an interaction sufficiently strong to produce the
Ne\`el
instability, but the validity of this approach at large doping
levels
is somewhat
questionable. In the Anderson scenario both spinons and holons
contribute to the resistivity, but it is often tacitly assumed
that the
scattering of holons dominates, while only spinons, which
effectively commute with the magnetic field, $B$,
appear in the temperature
variation of the Hall angle. 
Although commonly used to analyze experiments, this proposal has yet to be
verified by an explicit calculation of the transport
coefficients.\cite{pwa-private}

Coleman et al\cite{piers} have developed a
phenomenological transport theory based on the assumption that
the
dominant scattering in cuprates is mediated by an
interaction
which distinguishes quasiparticle wave functions with respect
to their charge conjugation symmetry. Such interactions are
absent in
ordinary FLs and hence this can be regarded as a NFL model,
although
in this case there is no spin-charge separation and hence the
model is more correctly categorized as an {\em unusual} FL. In
this
model quasiparticles with different symmetry have vastly
different
scattering rates, which in turn have different temperature
dependences, ultimately leading to the correct behavior of all
transport coefficients. Although the origin of the unusual
interaction has not
been identified, this is the only model to date which appears to
be
consistent with the recent experimental
results on optical reflectivity.\cite{sigmaxy-www}

The anomalous Hall effect can also be understood in terms of
the bipolaron theory, developed by Mott and
Alexandrov.\cite{mott}
Anomalous behavior in their model arises from Anderson
localization of bipolarons: at low temperatures bipolarons in a
random
potential lose their mobility, producing an enhanced value of
$R_H$.
At sufficiently high temperature, larger than the random
potential
binding energy,  the mobility is recovered, and
$R_H$ saturates, as seen in experiment. The bipolaron
model is also consistent
with a number of photoemission experiments.

A variety of FL-based models have also been utilized to
explain the anomalous transport in cuprates.
Clearly, in order to account for experiments, one must introduce
the {\em two} scattering rates without violating the FL ground
state.
This is done either by introducing a certain anisotropy
of scattering or by assuming a different mechanism of scattering
in the
presence of magnetic field.
Initially, the anomalous Hall effect was attributed to the
well-known
skew scattering present in many materials with strong spin-orbit
coupling;\cite{fiory} the anomalous temperature variation
of $\rho_{xy}$ arises from
anisotropic scattering of conduction electrons by local magnetic
moments. This approach naturally produces the observed
temperature
dependence of $\rho_{xy}$, provided the resistivity is
assumed linear in $T$. However, one runs into problems for
systems
with doping levels for which
the resistivity is not proportional to
temperature. Moreover, skew scattering should saturate with
increasing magnetic field, due to the complete polarization of
localized moments. Such saturation has not been observed,
although some
discrepancy from the linear in $B$ Hall resistivity does occur
in extremely high pulsed fields.\cite{bo}

Levin and Quader\cite{levin}
focused on a very important subgroup, the bilayer
cuprates. In their approach the anomalous behavior of the Hall
effect
is due to the multiband structure of bilayer cuprates. Their
main
assumption
is that one of the bands is very close to the Fermi level and
responsible for the anomalous resistivity, while the
other
is responsible for the Hall angle. The former is thermally
activated at higher temperatures in underdoped materials
so that the model can explain qualitatively
the behavior of resistivity in the underdoped bilayers.
Although their approach appears
reasonable, the striking similarity between the experimental
results on
single layer and bilayer materials, which we discuss in Sec.\
\ref{sec:experiment}, indicates that this approach may have
to be revised before it possesses general applicability.

Another class of models is based on anisotropic scattering,
which can arise from either  anomalous band structure in the
cuprates
or spin-fluctuation scattering.
Newns\cite{newns} has proposed a model
based on an anomalously large density of states near the Fermi
surface
(FS) for quasiparticles in the
vicinity of the so called Van Hove singularities: angle resolved
photoemission (ARPES) experiments in a number of samples show
extended flat band features near $k=(\pi,0)$ and symmetry
related points in the Brillouin 
zone, which give rise to highly anomalous
quasiparticle scattering. The two scattering
rates necessary to explain the measured transport properties
are those for quasiparticles near and far away from these
singularities. Although the model can produce a temperature
dependent Hall coefficient, $R_H\equiv \rho_{xy}$, it is unclear
whether it can simultaneously account for the anomalous behavior
of the resistivity.

We next consider models based on spin-fluctuation scattering.  
One of the first contributions has been provided by Carrington,
Cooper and their collaborators\cite{cooper} who used a phenomenological
model of the spin-fluctuations and a realistic FS and
calculated a number of transport coefficients in cuprates. They assumed
that large parts of the FS have an anomalous, linear in $T$
scattering rate, while smaller parts have a Fermi liquid,
quadratic in $T$ scattering rate.
Although the anomalous quasiparticle
scattering rate, assumed to be due to
spin-fluctuations, has been taken as an input to the theory, they
obtained a remarkable agreement not only with the Hall effect, but 
with the thermoelectric power as well. This pioneering work motivated a
number of authors to explore more particular choices of
spin-fluctuation spectra and their relationship with
transport.\cite{tmrice,lercher,SP}

Trugman\cite{trugman} calculated the band structure
renormalization in the presence of strong antiferromagnetic (AF)
fluctuations within the context of the
$t-t^\prime-J$ model and obtained a single particle band structure
which has the form of a precursor to a spin-density-wave (SDW)
state.
In this case  one finds flat regions in the band, somewhat
similar to
those mentioned above, which are
characterized by an energy scale which is
much smaller than the fermionic bandwidth.
In essence the size
of the Fermi energy is effectively reduced to an energy
$\Omega\ll t$,
and hence, just as in ordinary FLs, the resistivity deviates
from its usual $T^2$ dependence, while the Hall coefficient is
temperature dependent.
However we note that this anomalous temperature
dependence of the Hall effect
is primarily due to the strong anisotropy of
scattering rates
and densities of states in
different regions of the Brillouin zone.
Similar models were also provided by 
Kim et al\cite{kathy} and by Dagotto et al.\cite{dagotto}

Lercher and Wheatley\cite{lercher}
also studied a spin-fluctuation model of magnetotransport in
the cuprates. They concluded that
the behavior of several quantities of
interest is significantly altered depending on
whether one incorporates Umklapp processes. At high
temperatures they find
a highly $T$ dependent Hall coefficient, with a resistivity
approximately linear in $T$.
 As in the above cases, their
calculation is based on the anisotropy of scattering rates in
different regions of the Brillouin zone, although here the anisotropy
originates
in the specific structure of the effective interaction.

The nearly antiferromagnetic Fermi liquid (NAFL) model for transport in the
cuprates resembles the model studied by Lercher and Wheatley, as well as the
earlier Hubbard model calculations of the resistivity by Bulut, 
Scalapino and White,\cite{scalapino}
with one important difference:  the choice of the effective
interaction.  Although in both cases the effective quasiparticle interaction,
$V_{eff}$, assumed to be of magnetic origin, is
 a strong function of momentum transfer, and is
highly peaked near the antiferromagnetic wavevector ${\bf Q}$,
Lercher and Wheatley calculate $V_{eff}$ using an RPA formalism
with a
local restoring force, while in the NAFL model one takes the
momentum
and frequency dependence of the interaction to be that of
the spin-spin
response function which provides a quantitative account of the
NMR
experiments,\cite{BP} as discussed in the next
Section. While an RPA formalism can provide
insight into the physical
origin of the spin-fluctuations, it neglects changes in the effective
interaction brought about by the  spin-fermion vertex corrections,
which lead to a sequence of crossovers in the low frequency magnetic
behavior in
the normal state.\cite{c-p-s}  Thus it is not capable of
accounting for the rich morphology of the magnetic
properties seen in NMR experiments,\cite{charlie} a morphology which we shall
see is reflected in transport experiments.

Pioneering calculations of the resistivity of YBa$_2$Cu$_3$O$_7$ using a NAFL
model were carried out by Monthoux and Pines,\cite{MP}
who found, in a strong coupling (Eliashberg) calculation which incorporated
vertex corrections, that a good quantitative account of experiment could be
obtained using a spin-fluctuation spectrum taken from NMR experiments and a
coupling constant which yielded a superconducting transition at $90$K.  This
work provided an important bridge between the anomalous normal state properties
and the superconducting transition.  There remained the important question of
whether an NAFL model could also account for the anomalous Hall effect.  The
present authors addressed this question by using standard Boltzmann transport
theory to calculate both the resistivity, $\rho_{xx}$, and the Hall
conductivity, $\sigma_{xy}$ of YBa$_2$Cu$_3$O$_7$.\cite{SP}
  A direct (nonvariational) numerical solution of the BE showed that the
highly anisotropic quasiparticle scattering at different regions
of the Fermi surface (brought about by the highly anisotropic magnetic
quasiparticle interaction which characterizes a NAFL) could give rise
qualitatively, and in some cases quantitatively, to the measured anomalous
temperature dependence of the resistivity and Hall coefficient while yielding
the quadratic temperature dependence of the Hall
angle.\cite{SP,bps}  The present paper extends these calculations
in a number of ways.  First, we use perturbation theory to calculate, for a
broad range of doping and temperature regimes, the anisotropic quasiparticle
scattering rates around the Fermi surface; we then use these results to
calculate the longitudinal and Hall conductivities.  Our perturbation theory
approach not only makes explicit the difference in quasiparticle behavior at hot
spots (those regions of the FS for which the magnetic interaction is anomalously
large) and in cold regions (the remaining parts of the FS), but enables one to
examine in detail the changes in quasiparticle lifetimes in both regions brought
about by both the crossovers in the spin-fluctuation spectra measured in NMR
experiments and the changes in FS shape measured in experiments using angle
resolved photoemission techniques.  On taking the relevant averages of
quasiparticle lifetimes for the longitudinal and Hall conductivities, we are
then able to obtain simple approximate expressions for these quantities as a
function of doping and temperature which are in good agreement with experiment.

The substantial anisotropy of quasiparticle mean free paths found in the
perturbation theory approach led us to develop a phenomenological model for the
variation in mean free path as one goes around the FS, and to use this to
extract from experiments on $\sigma_{xx}$ and $\sigma_{xy}$ the behavior of the
quasiparticle lifetimes, $\tau_{\rm hot}$ and $\tau_{\rm cold}$, as one varies
temperature and doping level for the YBa$_2$Cu$_3$O$_{6+x}$ (123), 
La$_{2-x}$Sr$_x$CuO$_4$ (214), and Tl-doped systems.
Again, good agreement is found between the perturbation theory estimates and
the experimental results.

Guided in part by the perturbation theory results and our phenomenological
model, we then carry out direct numerical solutions of the BE for
representative members of the 123 and 214 systems, and for a highly simplified
model of a system, such as YBa$_2$Cu$_4$O$_8$ 
(124), in which the presence of chains brings about
considerable planar anisotropy.  For all three cases considered, good agreement
is found between theory and experiment.

The paper is organized as follows: in the next section we
provide the necessary background on the signatures of NAFL model
in NMR
experiments and define the relevant crossover temperatures.
In Sec.\
\ref{sec:theory} we analyze the properties of NAFLs using a
simple perturbative approach. In Sec.\
\ref{sec:experiment} we review experimental results found for the 123,
214, and overdoped Tl systems, and call attention to the regularities (or
lack thereof) which
$\rho_{xx}(T)$ and
$\rho_{xy}(T)$ exhibit as a function of doping and temperature.
We then present our phenomenological analysis of $\sigma_{xx}$ and
$\sigma_{xy}$ in terms of a highly anisotropic mean free path, and use this
analysis to extract from experiment the quasiparticle lifetimes in the hot and
cold regimes.
In Sec.\ \ref{sec:numerics}, we present
numerical calculations for a realistic set of band and
spin-fluctuation parameters, and compare our results to
experiment.  
Finally, in Sec.\ \ref{sec:conclusions} we summarize our conclusions.

\section{Background on Physical Properties of NAFLs}
\label{sec:background}
\typeout{background on \thepage}

In the NAFL description of the normal state properties of the
superconducting cuprates it is the magnetic interaction between
planar
quasiparticles which is responsible for the anomalous  spin and
charge
behavior. The magnetic properties of the system are specified by
a
dynamical spin-spin response function of fermionic origin,
$\chi({\bf
q},\omega)$, which near a peak at ${\bf Q}_i$, in the vicinity
of the commensurate wavevector,
${\bf Q}=(\pi,\pi)$, takes the form:
\begin{equation}
\chi(q, \omega) =  \sum_i \frac{\chi_Q}{1 + ({\bf q} - {\bf
Q}_i)^2 \xi^2 - i
\omega/\omega_{sf} -
\omega^2/\Delta^2}
\label{eq:mmp}
\end{equation}
Here $\chi_{Q_i}\gg \chi_0$ is the magnitude of the static spin
susceptibility at a (possibly) incommensurate
wavevector, ${\bf Q}_i$, near
${\bf Q}$, $\xi$ is the antiferromagnetic correlation
length, $\omega_{sf}$ specifies the low frequency relaxational
mode,
brought about by the near approach to antiferromagnetism, and
$\Delta$
allows for the possible presence of a spin gap at large
energies.
We have assumed a system of units in which the lattice spacing $a=1$.
The quasiparticle spectrum is assumed to take the tight-binding
form,
\begin{equation}
\epsilon_k=-2t( \cos k_x + \cos k_y) - 4t^\prime \cos k_x \cos
k_y-
2 t^{\prime \prime} (\cos (2k_x) + \cos (2k_y))
\label{eq:dispersion}
\end{equation}
where $t$, $t^\prime$ and $t^{\prime\prime}$ are the appropriate
hopping integrals, while the effective magnetic interaction
between the
planar quasiparticles is specified by
\begin{equation}
V_{eff}({\bf q},\omega)=g^2\chi({\bf
q},\omega).\label{eq:interaction}
\end{equation}
For a given system the parameters
$\chi_{Q_i}$, $\xi$, $\omega_{sf}$ which determine
$\chi({\bf q},\omega)$ are
taken from fits to NMR and INS (inelastic neutron scattering)
experiments, while the effective coupling constant, $g$, is
assumed to
be momentum independent for the wavevector near ${\bf Q}_i$. As
discussed by
Chubukov et al,\cite{c-p-s} in principle the effective
interaction,
Eq.\ (\ref{eq:interaction}), can be derived microscopically
starting
with, e.g., a one-band Hubbard model, but for a number of
reasons it
has not yet been possible to carry through such a derivation.

Since in practice the effective damping of the magnetic
excitations is
large, the term proportional to $\omega^2$ in the denominator of
Eq.\
(\ref{eq:mmp}) can be neglected, so that we may write:
\begin{equation}
\chi({\bf q}, \omega) =  \frac{\alpha
\omega_0}{\omega_{q} - i\omega}.
\label{eq:chi}
\end{equation}
where we have introduced the scale factor, $\alpha$, which
relates $\chi_{Q_i}$ to $\xi^2$,
\begin{mathletters}
\begin{equation}
\chi_{Q_i} = \alpha \xi^2,
\end{equation}
\begin{equation}
\omega_q = \omega_{sf} + \omega_0 ({\bf q}-{\bf Q}_i)^2.
\end{equation}
and
\begin{equation}
\omega_0 = \omega_{sf}\xi^2,
\end{equation}
\end{mathletters}
In the next Section 
we shall for the most part assume that the spin
fluctuation spectrum possesses only a single peak at ${\bf Q}$;
we consider,
in Sec.\ \ref{sec:numerics}, the effects of incommensuration on the
transport properties.

The effective interaction $V_{eff}$, Eq.\
(\ref{eq:interaction}),
has an obvious property: for
sufficiently large correlation lengths it is highly peaked for
momentum transfers in the vicinity of the antiferromagnetic
wavevector ${\bf Q}$. The importance of this fact is hard to
overestimate: if the FS of the system of fermions, defined by
the
quasiparticle dispersion (\ref{eq:dispersion}), is such that it
intersects the magnetic Brillouin zone Z (see Fig.\ \ref{fig:fs}), then
quasiparticles in the vicinity of these
intersection points on
the FS, often referred to as hot spots,\cite{tmrice}
 are much more strongly scattered by the spin-fluctuations than
those which are on other parts of the FS. This
is the main source of the anisotropy of quasiparticle behavior
in this model, since the temperature
variation of the quasiparticle
scattering rates at and far away from hot spots is
in general very different.
We shall return to
this point in the following Section.

Barzykin and Pines (BP) have utilized Eq.\ (\ref{eq:mmp}) in
their
analysis\cite{BP}
of the NMR results in HTS. Because their work, subsequently extended to
an analysis of INS results,\cite{zbp}
provides an important
insight into the morphology of the magnetic spectrum in these
materials, we briefly review their main conclusions.
BP find that for underdoped systems the low frequency  magnetic
behavior possesses three distinct temperature regimes, with a
crossover
at $T_{cr}$ from non-universal  mean field (MF) behavior with
dynamical
exponent $z=2$ to $z=1$ pseudoscaling (PS) behavior. For $T\geq
T_{cr}$,
$\omega_{sf}\sim 1/\xi^2$, and the product,
$\chi_{Q_i}\omega_{sf}\sim\omega_{sf}\xi^2=\omega_0$ is
independent of
temperature, while between $T_{cr}$ and $T_*$, it is
$\omega_{sf}\xi$
which is independent of temperature.
NMR experiments show that above $T_{cr}$, in the MF regime,
$\omega_{sf}$ and $\xi^2$
scale linearly with $T$, i.e., $\omega_{sf}=A+BT$, while between
$T_*$
and $T_{cr}$, the PS regime,
it is $\omega_{sf}$ and $\xi^{-1}$ which scale linearly
with $T$, albeit with a somewhat different slope and intercept
of $\omega_{sf}$ than
that found above $T_{cr}$.
 At $T_{cr}$, the temperature dependent uniform
susceptibility, $\chi_0(T)$, possesses a maximum, while $\xi\sim
2$, as
has now been verified experimentally for the
YBa$_2$Cu$_3$O$_{6.92}$,
YBa$_2$Cu$_4$O$_8$, and La$_{2-x}$Sr$_x$CuO$_4$
compounds.\cite{zbp}
Below $T_*$, in the
pseudogap (PG) regime, $\xi$ becomes independent of temperature
while
$\omega_{sf}$, after exhibiting a minimum near $T_*$, rapidly
increases as $1/T$ as $T$
decreases toward $T_c$.  
In addition to the change in the magnetic
fluctuation spectrum, the PS regime is characterized by a
strong temperature variation of the quasiparticle band structure,
resulting in a FS evolution,\cite{c-p-s} 
This evolution has non-trivial consequences for the
transport, as discussed in the next section.

From a magnetic perspective, the so called optimally doped
systems (e.g., YBa$_2$Cu$_3$O$_{6.93}$ and
La$_{1.85}$Sr$_{0.15}$CuO$_4$) are a special case
of the underdoped systems, in which $T_*$ is comparatively close
to
$T_c$. As may be seen in Table \ref{tab:crossovers}, $T_{cr}$
varies
rapidly with doping, and approaches $T_c$ as the system
approaches
the overdoped regime, while $T_*$, which varies relatively
weakly with
doping possesses a shallow maximum ($\sim 200$K in the 123
system, $\sim
100$K in the 214 system) as one varies the hole content in the
planes.

From this same magnetic perspective, overdoped cuprates are
defined as
those for which $T_{cr}<T_c$. For these systems, then, since
$\xi<2$, the
antiferromagnetic correlations are comparatively weak,
$\chi_0(T)$ is at most weakly temperature dependent, while
$\omega_{sf}\propto \xi^2$ follows the linear in $T$ behavior
found  in
the underdoped systems above $T_{cr}$. Examples of overdoped
systems
are the $T_c\sim 40$K Tl 2212 system and La$_{2-x}$Sr$_x$CuO$_4$
for
$x\geq 0.2$.

As we shall see, the crossovers seen in the low frequency
magnetic
behavior possess to a remarkable extent their charge
counterparts
in magnetotransport and optical experiments.

\section{Perturbation Theory Approach}
\label{sec:theory}
\typeout{Section theory on Page \thepage}

In this section we estimate the transport coefficients in
NAFLs using a simple perturbation theory. We calculate the
scattering
rates for
quasiparticles near the FS, assuming the effective interaction
$V_{eff}$,
Eq.\ (\ref{eq:interaction}),
and then use these to calculate the longitudinal and Hall
conductivities. We discuss the behavior of these quantities in
the relevant doping and temperature
regimes.

Within the Boltzmann-Bloch formalism the relaxation rate for a
particle near the FS is generally given by:\cite{ziman}
\begin{equation}
{1\over \tau_k}=\int {d^2{k}'\over (2\pi)^2} M({\bf k},{\bf
k}')
(1-\Phi_{k^\prime}/\Phi_k)\label{eq:coll}
\end{equation}
where $\Phi_k$ is the quasiparticle distribution function
displacement
obtained by solving the linearized
Boltzmann equation. To second order in the coupling constant
the scattering rate between points ${\bf k}$ and
${\bf k}^\prime$ is given by
\begin{equation}
M({\bf k},{\bf k}')=
2g^2 \, {\rm Im}
\chi({\bf k-k^\prime},\epsilon^\prime-\epsilon) \,
[n(\epsilon^\prime-\epsilon) + f
(\epsilon^\prime)]\label{eq:scatkq}
\end{equation}
where $\epsilon\equiv \epsilon ({\bf k})$,
$\epsilon^\prime\equiv
\epsilon ({\bf
k}^\prime)$,
$n(\epsilon)$,  and $f(\epsilon)$ are the Bose and Fermi
distribution functions respectively.
Note that Eq.\ (\ref{eq:scatkq}) is obtained by summing the
Matsubara
frequencies and assuming a retarded scattering
potential.\cite{rickayzen}
No lifetime effects are included.
One can, of course, easily  add the  self-energy corrections to
the
bare band spectrum $\epsilon_k$, but the life-time effects are
considerably harder to include and in
principle
should be obtained self-consistently.
In the relaxation time approximation
one replaces $\Phi_k$, the quasiparticle
distribution displacement,
by $\varphi(T) ({\bf v\cdot n})$, where $\varphi$ is a
function of temperature independent of ${\bf k}$.
The assumption that $\varphi$ does not vary greatly in the
Brillouin zone
is most certainly invalid for HTS,\cite{bps} given the strong
momentum dependence of the effective interaction $V_{eff}$, Eq.\
(\ref{eq:interaction}). However, it is still quite reasonable,
as long as the  temperature dependence of $\varphi$ is the
same in each of the hot spots, which is almost always the case.
Since the inclusion of $({\bf v\cdot n})$ produces only a
{\em geometric} factor (vertex correction), which is
large mostly for small momentum
transfer, 
and no additional temperature dependence, we
can approximate the relaxation rate by the scattering rate,
which, in the  second order Born approximation reads,
\begin{equation}
{1\over \tau_k} = \int {d^2{k}'\over (2\pi)^2}
M({\bf k},{\bf k}')\label{eq:tau}.
\end{equation}
One can make a change of variables in the integral in Eq.\
(\ref{eq:tau}):
\begin{equation}
\int d^2{k}'\rightarrow \int d\epsilon' \int {ds' \over \vert
{\bf v}\vert}\label{eq:varchange}
\end{equation}
and solve the integral over $\epsilon'$ analytically for
${\bf k}$ near the FS, ($\epsilon\approx 0$), using:
\begin{equation}
\int dx\, n(-x) \,{-x\over \Omega^2 + x^2}\, f(x)={\pi\over
2\Omega} +
\pi
\sum_n {(-1)^n\over \Omega + n\pi}= {\pi\over 2\Omega} - h\left(
1+{\Omega\over \pi}\right)\label{eq:integral}
\end{equation}
where\begin{equation}
h(x)={1\over 2}
\left[ \psi\left({x+1\over 2}\right) - \psi \left({x\over
2}\right)
\right]
\end{equation}
which has the following limiting behavior:
\begin{eqnarray}
&h(1+x)\approx \ln 2 - {\pi\over 2} x\qquad &x\rightarrow 0\\
&h(x)\approx {1\over 2} {1\over x+1}\qquad &x\rightarrow \infty
\end{eqnarray}
Thus the integral in Eq.\ (\ref{eq:integral}) can be
approximated by:
\begin{equation}
{\pi\over 2}\left({1 \over \Omega} - {1\over \Omega +
\pi}\right)
\end{equation}
and the scattering rate, Eq.\ (\ref{eq:tau}), is given by:
\begin{equation}
{1\over \tau_k} = {\alpha
g^2\omega_0\over 4}\int_{FS} {dk'} \tilde{M}
(k,k^\prime)\label{eq:scatt}
\end{equation}
where
\begin{equation}
\tilde{M}(k,k^\prime)={1\over |{\bf v}|}\,
{T^2\over \omega_{kk'} (\omega_{kk'}+\pi T)}.\label{eq:tildem}
\end{equation}
Note that in a typical FL one has an identical expression;
however
$\omega_{k,k^\prime}$ is large,
temperature independent and
only weakly dependent on ${\bf k,k^\prime}$. Then
for $T\ll \omega_{kk^\prime}$ the scattering rate has the usual
$T^2$ temperature dependence which ceases only
when $T\sim\omega_{k,k^\prime}$. In a realistic metal,
however, this energy scale can be considerably reduced for
certain
${\bf k}$
due to the presence of band singularities at these points.
In a NAFL, however, $\omega_{k,k^\prime}$ is a strong function
of
momentum
transfer ${\bf q}$, i.e., $\omega_{k,k^\prime}$ has a minimum
for ${\bf q}={\bf Q}$. Thus, the scattering rate $\tilde{M}$ is
maximized
for points ${\bf k}$ and ${\bf k}^\prime$ 
on the FS such that ${\bf q}={\bf 
k-k^\prime}={\bf Q}$.

We now calculate
the scattering rate for a quasiparticle near the FS as a
function of
$T$ and $\Delta k$, where $\Delta k$ is the distance to
a nearby hot spot along the FS. For the commensurate
case and a large FS
the hot spots are usually distributed along the FS in such
a way that one can write,
for ${\bf k-k^\prime}\approx {\bf Q}$, that $({\bf
k-k^\prime-Q})^2\approx
(\Delta k)^2 + (\Delta k^\prime)^2$, where $\Delta k$ and
$\Delta k^\prime$
are the displacements of ${\bf k}$ and ${\bf k}^\prime$ from
the nearby hot spots (see Fig.\ \ref{fig:fs}) (the adjoint hot spots
 correspond to $\Delta k,\Delta k^\prime = 0$).
This basically means that there is no strong nesting at ${\bf q}
\approx {\bf Q}$, and the FS and its shadow FS, obtained by translating 
the FS by ${\bf Q}$, intercept at a high angle $\Omega\sim \pi/2$.
Then one easily finds: 
\begin{equation}
{1\over \tau_k} = {\alpha g^2 T \sqrt{\omega_0}\over 4 v_f}
\left({1\over \sqrt{\omega_{sf}+\omega_0(\Delta k)^2}}
- {1\over \sqrt{\omega_{sf}+\pi T+\omega_0(\Delta k)^2}}\right)
\label{eq:1overtau}
\end{equation}
where we have assumed that the integration over $\Delta
k^\prime$ in
Eq.\ (\ref{eq:tildem})
can be extended to infinity. Eq.\ (\ref{eq:1overtau}) can be
rewritten as:
\begin{equation}
\tau_k = {4 v_f Q_k\over \alpha \pi g^2 T^2\sqrt{\omega_0} }
\label{eq:scatt_time}
\end{equation}
where
\begin{eqnarray}
Q_k&=& \sqrt{\omega_{sf}(1 + (\xi\Delta k)^2)}
\sqrt{\omega_{sf}(1 + (\xi\Delta k)^2) + \pi T} \times
\nonumber\\
&&\left( \sqrt{\omega_{sf}(1 + (\xi\Delta k)^2)} + 
\sqrt{\omega_{sf}(1+(\xi\Delta k)^2)  + \pi T}
\right)
\label{eq:q_k}
\end{eqnarray}
Note that this result is easily generalized to arbitrary $\Omega$ by 
replacing $\Delta k$ with $(\Delta k)\sin\Omega$.

Before proceeding with the transport coefficients
it is instructive to calculate the
scattering rates for different values of $\Delta k$.
Eq.\ (\ref{eq:q_k}) suggests that there are primarily two
regions of
interest: we call a region of momenta $k$ hot if
\begin{equation}
\Delta k \leq 1/\xi
\end{equation}
and we call a region cold if
\begin{equation}
\Delta k \geq 1/\xi\label{eq:hotcold}
\end{equation}
Clearly the relaxation times in these two regions are different.
In the hot region one has
\begin{equation}
Q_k\approx \sqrt{\omega_{sf}}\sqrt{\omega_{sf} + \pi T}
\,(\sqrt{\omega_{sf}} + \sqrt{\omega_{sf} + \pi T})
\end{equation}
The fits to NMR experiments of Barzykin and Pines,\cite{BP} show
that in all HTS one can assume for practically all temperatures of 
interest that, to a good approximation,
\begin{equation}
\omega_{sf}\ll \pi T
\end{equation}
and hence in the hot region we have
\begin{equation}
{1\over \tau_k} \approx {\alpha g^2 T\xi\over 4 v_F}
\label{eq:lametauhot}.
\end{equation}
On the other hand, in the limit $\xi\gg 1$, in the cold region
one has
\begin{equation}
Q_k\approx \sqrt{\omega_{0}(\Delta k)^2} \sqrt{\pi T + \omega_0
(\Delta k)^2} (\sqrt{\omega_{0}(\Delta k)^2} + \sqrt{\pi T +
\omega_0
(\Delta k)^2} ).
\end{equation}
Here we distinguish two different cases, depending on the value
of $\Delta k$, yielding:
\begin{eqnarray}
{1\over \tau_k}&={\alpha g^2 T\over 4v_f (\Delta k)}\qquad &\pi
T\gg\omega_0 (\Delta k)^2\label{eq:lamecoldsmallk}\\
{1\over \tau_k}&={\alpha \pi g^2 T^2
\over 8v_f (\Delta k)^3 \omega_0}\qquad &\pi T\ll
\omega_0 (\Delta k)^2\label{eq:lamecoldlargek}
\end{eqnarray}

Obviously, the scattering rates depend on the particular
temperature
regime for the magnetic fluctuations, through the temperature
dependence of $\omega_{sf}$ and
$\omega_0$. In the $z=2$ MF regime
$\omega_0=\omega_{sf}\xi^2=const$ and
$\omega_{sf}=A+BT$, and therefore
it is easy to verify (see Eq.\ (\ref{eq:q_k})) that
the behavior of the scattering rate for any $\Delta k$ can be
obtained
from the $\Delta k=0$ case by
 replacing $A$ in $\omega_{sf}$
	with $A + \omega_0(\Delta k)^2$. In other words, every
point on the FS has a similar shape of $1/\tau_k$;
the hot region is enlarged with increasing temperature, as can
be verified from Eq.\ (\ref{eq:hotcold}).

In the $z=1$ PS regime the situation is somewhat different, due to
the
temperature dependence of $\omega_0$, $\omega_0\sim 1/T$. First,
from
Eq.\ (\ref{eq:lametauhot}) it is clear that the scattering rate
due to
spin-fluctuation scattering is roughly temperature independent
in the hot region. The scattering rate in
Eq.\ (\ref{eq:lamecoldsmallk}), valid for $\xi\Delta k\ll
\sqrt{\pi
T/\omega_{sf}}$, is applicable in this case as well. Finally,
far away
from hot spots, Eq.\ (\ref{eq:lamecoldlargek}), the scattering
rate is
proportional to $T^3$. Note that the condition $\omega_0 (\Delta
k)^2\gg \pi T$ is satisfied for significant portions of the FS at
very low temperatures.
The scattering rates as functions of temperature for several
different
values of $\Delta k$, in MF and PS regimes,
are given in Figs.\ \ref{fig:ratesMF}a and \ref{fig:ratesPS}a.

In the PG case the situation is even more complicated and
somewhat unclear, due to the FS evolution mentioned in Sec.\
\ref{sec:background}. Here $\Delta k$ does not correspond to a fixed
point on the FS, since the FS itself migrates with decreasing
temperature. Nevertheless, provided that the FS is still
relatively
large, the scattering rates are nominally the same as in the PS
case, since $\omega_0=\omega_{sf}\xi_c^2\sim 1/T$,
where $\xi_c\sim const$ is the correlation
length at $T_c$, and $\omega_{sf}\sim 1/T$. However,
the scattering rate at  a particular point in the Brillouin zone
is non-trivial and we do not present it here.


So far we have neglected the fact that in deriving
Eq.\ (\ref{eq:1overtau}) the integral over $\Delta k^\prime$ was
performed from $(-\infty, \infty)$.
This is quite reasonable provided that the
system is far from the PG regime. However, as mentioned in
Section
\ref{sec:background}, as the temperature is
lowered toward $T_*$ the quasiparticle spectral
weight is removed from the hot spots
due to strong spin-fermion vertex corrections and at $T\sim T_*$
the
system
actually starts to lose parts of the FS. Hence, one must assume
a
finite limit of integration over $\Delta k^\prime$, yielding a
somewhat
more cumbersome form of Eq.\ (\ref{eq:1overtau}) which we do not
quote
here. However, it is straightforward to show that although this
does not affect the cold regions very much, the scattering rate
in
the hot region is considerably modified, as shown in
Figs.\ \ref{fig:ratesMF}b and \ref{fig:ratesPS}b.
 Although somewhat surprising, this result is natural in view of
the FS
modulation: the
hot spots migrate with decreasing temperature and the effective
size
of the FS
is reduced, thereby reducing  the scattering of quasiparticles
in the magnetic
channel. More importantly, since the FS is small, i.e., $\Delta
k_{max}$ is
very small, the scattering is anomalous around most of the FS,
as
can easily be verified from Eq.\ (\ref{eq:q_k}). We return to
this
point
shortly.

The low field longitudinal and Hall conductivities  are
calculated according to:
\begin{eqnarray}
\sigma_{xx}&=&-2e^2\sum_k \,({\bf v_k\cdot n})^2\,\tau_k \,
\left({\partial
f_0\over
\partial\epsilon}\right)\label{eq:sigmaxx0}\\
\sigma_{xy}&=&-2e^3\sum_k \,({\bf v_k\cdot n_\perp\tau_k})\,
{\bf v}_k\times {\bf B}\cdot \nabla
({\bf v_k}\cdot {\bf n}\tau_k) \left({\partial f_0\over
\partial\epsilon}\right)\label{eq:sigmaxy0}
\end{eqnarray}
where ${\bf n}$ and ${\bf n}_\perp$ are
unity vectors perpendicular to the applied electrical field $E$,
and ${\bf B}$ is the applied transverse magnetic field.

We start with the resistivity.
At relatively low temperatures, $T\ll E_f$, where $E_f$ is the
Fermi energy,
one can perform the change of variables (\ref{eq:varchange}) and
integrate over $\epsilon$. Assuming that the FS is four-fold
symmetric (tetragonal symmetry), the conductivity is given by:
\begin{equation}
\sigma_{xx}={e^2\over 4\pi^2} \int_{FS} dk\, \tau_k v_f
\end{equation}
Assuming that the Fermi velocity does not vary appreciably
around the FS, we find that the conductivity is
roughly proportional to the average of the scattering time
(\ref{eq:scatt_time}) around the FS. From Eq.\
(\ref{eq:scatt_time}) it
is clear that the main contribution
to the conductivity comes from the cold regions of
the FS where $\tau_k$ is the largest, while
the hot spots ($\Delta k\sim0$) contribute relatively
little. On averaging Eq.\ (\ref{eq:q_k}) we obtain
a rather cumbersome formula for the resistivity, which, to
within
logarithmic corrections, can be approximated by:
\begin{equation}
\rho_{xx} \approx {\alpha g^2\pi \over v_f}{T^2  \over T_0 + T}
{1\over
(\Delta k_{max})^2} \label{eq:rhoxx}
\end{equation}
where $T_0$ is a crossover temperature given by
\begin{equation}
T_0\approx \omega_0 (\Delta k_{max})^2/2\pi\equiv (\xi \Delta
k_{max})^2
\omega_{sf} /2\pi.
\label{eq:t0}
\end{equation}
If we define cold spots as these points on the FS which are maximally 
distant from a hot spot, then $\Delta k_{max}$ is the distance between 
a cold spot and the nearest hot spot. As may be seen in Fig.\ 
\ref{fig:fs}, the size of the $\Delta k_{max}$ is related to the 
departure of the FS from the magnetic  Brillouin zone boundary. For 
this FS and any large FS, such as is found by ARPES measurements for 
YBa$_2$Cu$_3$O$_7$ and related compounds, $\Delta k_{max}\sim 1$.
On the other hand, for a profoundly small doping level, or in the 
presence of substantive incommensuration (which increases the number of 
hot spots and acts to reduce significantly the distance between hot and 
cold spots), one finds $\Delta k_{max}\leq 0.3$.

It is evident from Eq.\ (\ref{eq:rhoxx}) that in the absence of any 
evolution of the Fermi surface with temperature, $\rho_{xx}$ will be 
linear in temperature as long as $T$ is appreciably greater than $T_0$.
In the MF regime, $T_0$ is independent of
temperature, while in the PS regime it increases as $\xi (\sim
1/T)$ as $T$
decreases. If $\Delta k_{max}$ and $\xi(T_*)$ are small
enough, then
$T_0\leq T_*$, and the resistivity will be linear in $T$ down to
$T_*$.
The exact condition for this extended linear in $T$ behavior
reads $(\Delta
k_{max})^2 \xi (T_*)\leq (T_*/100)K (\hat{c}/53{\rm meV})$,
where
$\hat{c}$ is the pseudoscaling spin velocity, $\omega_{sf}/\xi$.
Under these circumstances, any change in $\rho_{xx}$ at $T_{cr}$
will be
marginal, since it can only arise from changes
in the {\em hot} region contribution to $\rho_{xx}$.
The linearity in $T$ continues above $T_{cr}$
until $T\sim \omega_0$, an experimentally almost
unreachable temperature.
 Moreover, we have assumed that the only relevant
scattering is in the magnetic channel.
Strictly speaking, in addition to the contribution proportional
to
$\chi$ defined above one must  add the FL contribution,
$\chi_{FL}$,  present in all FLs,
to the effective interaction
$V_{eff}$  The latter is likely
 to dominate at these temperatures. Without this term at
$T>\omega_0$
one then has $\rho_{xx}\propto T^{1/2}$,  provided
$\omega_{sf}\propto T$.
 On the other hand, from Eq.\ (\ref{eq:q_k}) it is readily
 shown that if $\omega_{sf}\approx
const$ then $\rho_{xx}\propto T^2$. We shall return to this
point when
discussing overdoped samples.

What happens at $T_*$? As mentioned in
Sec. \ref{sec:background}, in the vicinity of $T_*$ 
the quasiparticle band structure begins to acquire a considerable
temperature dependence, implying a FS evolution as well.
Recently, Chubukov, Morr and Shakhnovich\cite{chub-fs}
studied the FS evolution as a function of the coupling 
strength $g$ at $T=0$ and 
found that with increasing spin-fermion interaction the FS acquires 
features of a preformed spin-density-wave state,
as depicted in Fig.\ \ref{fig:fsevolution}.
Alternatively one can show\cite{c-p-s} that a similar  
evolution occurs at fixed $g$, but as a function of temperature,
associated with a relatively strong 
temperature dependence of the spin-fluctuation damping.\cite{c-p-s}
As a result, 
the FS appears to get closer to the magnetic Brillouin zone 
boundary, i.e., the effective size of the FS decreases, with
decreasing temperature.
Hence when calculating transport coefficients one should
assume that $\Delta k_{max}$ acquires a temperature dependence, 
the precise form of which is not well known at present. 
Below $T_*$ one finds that $\omega_0$ retains its
$1/T$ overall temperature dependence
although it changes its
slope: $\xi$ becomes constant, while $\omega_{sf}$ increases as
$T^{-1}$.
Whether one gets a crossover at $T_0$
in the PG regime will then depend sensitively on the interplay
between $\Delta k_{max}(T)$ and $\omega_{sf}(T)$. If a
crossover to
$\rho_{xx}\propto T^2/T_0$ occurs, the resistivity
will then decrease
faster than $T^2$, due to the temperature dependence of $T_0$.

Let us examine the crossover in the vicinity of $T_*$ in more detail.
The above mentioned FS evolution usually closes channels for
scattering as the temperature is lowered below $T_*$.
New hot spots are developed and a smaller portion of the FS
is now {\em cold}. At the same time $\omega_0$ is strongly
temperature dependent in both PS and PG regimes. Assuming that
$T\gg T_0(T)$ above $T_*$ the reduction in size of the cold region
implies
that a  large part of the FS has a scattering rate
such as those shown in Fig.\ \ref{fig:ratesPS} for small values
of $\Delta k$, where $1/\tau$ is roughly independent of $T$. At a
certain temperature $T_*^r$ the FS is effectively so
small, that the quasiparticles around most of
the FS exhibit precisely these (almost constant in $T$)
scattering
rates; the resistivity thus acquires a finite negative curvature,
i.e.,
$d^2 \rho_{xx}/dT^2<0$. A signature of this should be
seen in the $T$ variation of the scattering rate at a hot spot
$\Delta
k\sim 0$. In the next Section we shall see that this is indeed
the
case. It is important to realize that this can
occur only when $\omega_0$ is strongly
temperature dependent; hence $T_*^r$ is never larger than
$T_{cr}$, where $\omega_0=const$, although in principle it can
be comparable to $T_{cr}$.
A rapid temperature dependence of $\omega_0$
eventually leads to a Fermi liquid like behavior at
$T\ll T_*^r$, regardless of the precise size of the FS.
Note that perturbation theory is only marginally applicable
in the PG regime, since one must include the strong
vertex corrections at or near
hot spots, although the fact that the conductivity
is dominated by regions away
from hot spots, where the above corrections are not as
pronounced,
allows us to make this qualitative argument.
We note that the
present theory is somewhat equivalent to the work of Quader et
al,\cite{levin}
although in their case the scattering channels
which open up at $T_*^r$ are
attributed to the multiband structure in bilayer cuprates and
the
proximity of one of the bands to the Fermi level.

Overdoped materials  exhibit only $z=2$ scaling and in general
$\omega_{sf}$ and $\xi$ show very little $T$ dependence. 
Quite generally in overdoped samples $\omega_{sf}=A+BT$ 
displays a much larger
value of $A$, associated with the strong fermionic damping of
the spin
fluctuations, found at high doping levels. Hence, even at a
hot spot
($\Delta k=0$) the
scattering rate $1/\tau_k$ is only weakly anomalous, i.e.,
$1/\tau_k\sim T$ or
even $T^2$.
Therefore in the overdoped samples one encounters a crossover
from
quadratic
in $T$ behavior of the resistivity to linear in $T$ behavior
above
$T_0$, even though the amount of the anomaly in the hot regions
is weak. However, there are extreme cases, such as
the  overdoped samples of 214 ($x\geq 0.25$), where the
anomalous
behavior is so weak that the resistivity is never linear
in $T$.
Moreover, in overdoped materials both the
large values of $\omega_{sf}$ and the
small values  of $\xi$ imply that the anisotropy of
scattering rates in overdoped materials is seriously reduced (see Eq.\
(\ref{eq:q_k}), producing
only a weakly temperature dependent Hall coefficient.

We now turn to the Hall effect.
Here the situation is, in principle,
a lot trickier, since the Hall coefficient is a
non-trivial function of temperature, due to
the strong anisotropy of the mean free path, $\ell\equiv \tau_k
v_k$,
as readily observed from the definition
of $\sigma_{xy}$ (Eq.\ (\ref{eq:sigmaxy0})).
However, in practice complications occur only if
$\ell$ has a very peculiar periodicity as $k$ goes around the
FS. For the FSs seen in ARPES measurements on a number of samples
near optimal doping one can safely assume:
\begin{equation}
\sigma_{xy} = {e^3 B\over 4\pi^2} \int_{FS} dk\,\ell_k^2
\label{eq:sigmaxyl}
\end{equation}
where we have performed an integration over $\epsilon$ in Eq.\
(\ref{eq:sigmaxy0}) in the usual manner.
Again, assuming that $v_f$ does not vary appreciably around the
FS, one
concludes that
$\sigma_{xy}$ is roughly proportional
to the average of $\tau_k^2$. Clearly, the leading contribution
to this
quantity comes from the regions of the FS where
$\Delta k\sim \Delta k_{max}$ (cold regions);
the hot regions contribute very
little. On averaging $\tau_k^2$ over $\Delta k$ (Eq.\
(\ref{eq:q_k})),
we make the same approximations as before, i.e., $\omega_{sf}\ll
\pi T$
and $\Delta k\,\xi \gg 1$ in the cold regions of interest. The
result
for $\sigma_{xy}$ is again cumbersome, but can be simplified by
expanding $\sigma_{xy}T^4$ in $T$, leading to a relatively
simple
formula:
\begin{equation}
\sigma_{xy}\approx {e^3 B v_f^2\over 4 \alpha^2 g^4\pi^2}
{(\Delta k_{max})^3\over T^4} \left(3T^2_0 + 6T_0 T +
T^2/\sqrt{2}\right)
\label{eq:sigmaxyemp}
\end{equation}
where $T_0$ is defined in Eq.\ (\ref{eq:t0}). Eq.\
(\ref{eq:sigmaxyemp}) is valid over an extended temperature
range.
Clearly, through the variations in $T_0$ and $\Delta k_{max}$,
it leads to a rather complex morphology of the Hall effect.

Our result,  Eq.\ (\ref{eq:sigmaxyemp}), suggests that
there are three temperature regimes for
the Hall effect in NAFLs: In the low temperature regime
$\sigma_{xy}\propto T_0^2/T^4$, in the high temperature regime 
$\sigma_{xy}\propto 1/T^2$, while at the intermediate temperatures
$\sigma_{xy}\propto T_0/T^3$. The exact values of the 
crossover temperatures 
depend greatly on the details of the band structure.
In what follows we shall assume, for the purpose of simplicity, that 
the crossover between the low and intermediate temperature regimes occurs 
at $T_0$ and that the crossover between intermediate and high 
temperatures occurs at $2\pi T_0$, where $T_0$, given by Eq.\
(\ref{eq:t0}),  is the crossover 
temperature observed in the resistivity;
the estimate of other quantities,
such as the Hall coefficient, is then rather convenient.
Our results for the transport coefficients are given in Table
\ref{tab:coefs}.
Note that for many systems away from PG regime one finds $\cot
\Theta_H\sim T^2$, due to the relatively large value of the
second
crossover temperature $T_2$.

Our discussion of the resistivity in underdoped cuprates
is easily extended to $\sigma_{xy}$:
in the MF regime $T_0$ is constant and hence
\begin{mathletters}
\begin{equation}
\sigma_{xy}\sim T^{-3}
\end{equation}
\begin{equation}
\cot \Theta_H\sim T^2
\end{equation}
\begin{equation}
\rho_{xy}\sim 1/T
\end{equation}
\end{mathletters}
for temperatures well below $\omega_0 (\Delta k_{max})^2$.
At somewhat higher temperatures the second crossover in
$\sigma_{xy}$
is approached so that both $\rho_{xy}$ and $\cot \Theta_H$ deviate
from this behavior.
In the PS regime $\sigma_{xy}\propto  T_0/T^{3}$, decreases
faster than
$1/T^{-3}$, since $T_0$ decreases with temperature. As $T$
approaches $T_{cr}$ one finds $\sigma_{xy}\sim T^{-3}$ even in this
regime.
Note that in this case the Hall coefficient varies faster
than $1/T$. Finally, due to the variation of $T_0$, the
second crossover can be approached, in which case $\rho_{xy}$ becomes
only weakly temperature dependent. We emphasize once more 
that in both MF and PS regimes the results depend a great deal on
details of the band structure. Moreover, the presence of impurities
makes a non-trivial effect on all of the quantities of interest.

As before the PG regime is even more complex due to the very
rapidly varying  $\omega_0$ and the FS evolution, which alters
$\Delta k_{max}$.
Although it is not clear what the temperature dependence of
$T_0$ is in
this case, due to the fact that at present one does not know
the temperature variation of the band structure in detail, we
can
still  make certain arguments.
 Near $T_*$ the value of $T_0$ changes rapidly and one can
easily switch between several regimes, due to the sensitivity of
the Hall effect to the variation in the 
band structure. For example, one can jump from
the first (low $T$) regime to the third, without noticing a
substantial
difference in, e.g.,  the Hall coefficient, since in both cases
the
quantity is roughly constant, while $\sigma_{xx}$ and
$\sigma_{xy}$
both experience non-trivial changes. 
Finally, at very low temperatures one finds
$\rho_{xy}$ approximately constant and $\cot\Theta_H$ varying
faster
than $T^2$.

The NAFL model suggests a very complex morphology of the
transport
coefficients in HTSs. We explore the extent to which this
morphology is
experimentally supported in the next Section.

\section{Analysis of Transport Measurements}
\label{sec:experiment}
\typeout{paramfs on Page \thepage}

In this section we  review briefly the in-plane normal state
transport
properties of the cuprate superconductors, with particular
attention to
the extent to which the doping and temperature dependence of the
planar
resistivity, $\rho_{xx}$, and the Hall conductivity,
$\sigma_{xy}$,
reflect the magnetic crossovers discussed in Sec.\
\ref{sec:background}
and the perturbation theoretic estimates of scattering rates
given in
Sec.\ \ref{sec:theory}. We then use a phenomenological model
which
incorporates the expected highly anisotropic quasiparticle mean
free
path
to deduce from experiments the detailed behavior of the
quasiparticle
scattering rates in both the hot and cold regions of the Fermi
surface.
We focus  on the results obtained for the bilayer 123 system (of
which
YBa$_2$Cu$_4$O$_8$, whose behavior corresponds closely to
that
found in YBa$_2$Cu$_3$O$_{6.68}$, is an ``honorary'' member), the
single layer La$_{2-x}$Sr$_x$CuO$_4$ (214) system, and as
representative of
overdoped materials,
the 15K single layer Tl 2201 material.

It is often stated that optimally doped materials, such as
La$_{1.85}$Sr$_{0.15}$CuO$_4$ and YBa$_2$Cu$_3$O$_{6.93}$
possess a
resistivity which is linear in $T$ over a wide temperature
region
which
extends down to $T_c$, However, close examination of single
crystal
data
shows that as $T$ approaches $T_*$, departures from linearity
occur in
these materials, departures which are in fact characteristic of
all
underdoped materials, but which become more pronounced as one
reduces
the
doping level below the optimum level. Quite generally,
$\rho_{xx}$ in
underdoped cuprates exhibits a drop (below linear behavior) at a
temperature, $T_*^r$, which for many systems is not far from the
temperature, $T_*$, at which the low frequency magnetic behavior
crosses
over from the PS to the PG regime. On the other hand in
overdoped
materials, $\rho_{xx}$ displays an upturn (from linear behavior)
with
decreasing $T$, an upturn which we attribute to the
comparatively weak
anomalous scattering in the vicinity of the hot spots we have
discussed
in Sec.\ \ref{sec:theory}. Thus one encounters qualitatively
different
departures from linearity, depending on whether one is
describing a
system which is overdoped or underdoped (using the magnetic
classification
proposed by Barzykin and Pines). Some of the experimental
results which
have led us to this conclusion are given in Fig.\
\ref{fig:rhotrho0}.
We note that the transport results presented in Fig.\
\ref{fig:rhotrho0}a
provide additional support for the assignment, based on analysis
of NMR
data,\cite{BP} of  ``optimally-doped'' YBa$_2$Cu$_3$O$_{6.92}$
material to the underdoped
sector.

To what extent does the crossover at $T_*^r$ correspond to the
magnetic
crossover at $T_*$? As may be seen in Fig.\ \ref{fig:rhotrho0}a
and
Fig.\ \ref{fig:rhotstartcr}, for YBa$_2$Cu$_3$O$_7$ and 
YBa$_2$Cu$_3$O$_{6.63}$ and
124, within
the present experimental uncertainties, the two crossover
temperatures
agree. However, the perturbation theory analysis of Sec.\
\ref{sec:theory}
suggests that this need not be the case in general, since if the
FS is
sufficiently far from the magnetic Brillouin zone boundary,
$T_*^r$ can
be considerably higher than $T_*$; this appears to be the case
for
the
underdoped members of the 214 family (see Fig.\
\ref{fig:rhotrho0}b for
La$_{1.9}$Sr$_{0.1}$CuO$_4$ and Table
\ref{tab:crossovers}). On the other hand, to our knowledge there
are no samples for which $T_*^r$ is appreciably lower than
$T_*$.

We have called attention, in Sec.\ \ref{sec:theory}, to the
possibility
of a crossover  to almost Fermi liquid like behavior
for $\rho_{xx}$. This occurs when $T<T_0(T)$ (see Eq.\
(\ref{eq:t0})).
This crossover is found in most, if not all, underdoped
materials, at sufficiently low temperatures, where $\rho_{xx}$
displays
a finite positive curvature above linearity,
$\rho_{xx}=\rho_{xx}(0) +
\beta T^b$, with $b$ larger than 1. Indeed, as may be seen in
Fig.\
\ref{fig:rhobelowtfl}, reasonable fits to the data may be
obtained with
$b\sim 2$. Note that here $\rho_{xx}(0)$ is the
actual residual resistivity, due presumably to disorder in a
sample.

In the overdoped materials one also
observes a crossover from $\rho= {\rho}_{xx}(0)+ \beta T^2$ to
linear in $T$ behavior at higher temperatures, as already
illustrated
by the Tl 2201 result in Fig.\ \ref{fig:rhotrho0}a. We emphasize
once again
that the physics of this crossover is different than in
underdoped materials, as  it arises from the fact that the the
anomaly
of scattering at hot spots is rather
weak in these compounds, rather than
in the pseudogap effects found in the underdoped cuprates.
It is important to notice,
however, that there is no second crossover, i.e., $T_*^r$
has not been observed.
Moreover, in some overdoped materials, such as
La$_{2-x}$Sr$_{x}$CuO$_4$
at the doping level $x=0.3$, the resistivity is never linear in
$T$, i.e., $\rho\approx T^{3/2}$
over the entire $T$ range.\cite{batlogg}

The morphology of the resistivity is still more complex
in materials with
CuO chains. As may be seen in Fig.\ \ref{fig:bucherrho}, where
we show
how $\rho_{aa}$ and $\rho_{bb}$ deviate from linearity in the
124 compound, chains lead to both a different magnitude
and a different temperature dependence for $\rho_{aa}$ and
$\rho_{bb}$ (see Ref.\ \onlinecite{bucher}).
Qualitatively similar results are found
for optimally doped 123,\cite{donbook} although the level
of a-b plane anisotropy is considerably less prominent.
On comparing Figs.\ \ref{fig:rhotrho0} and \ref{fig:bucherrho}
it may
be seen that the temperature dependence of
$\rho_{aa}$ and $\rho_{bb}$ in 124 resembles that found in the
underdoped and overdoped cuprates respectively. A detailed
explanation of this unique behavior in terms of NAFL theory is
given in  Section\ \ref{sec:numerics}.

We consider next the Hall effect in a transverse magnetic field.
In ordinary FLs one finds a  Hall conductivity which is roughly
proportional to the
square of the longitudinal conductivity,
($\sigma_{xy}\sim\sigma_{xx}\sigma_{yy}$),
and as a consequence the Hall resistivity, $\rho_{xy}=
\sigma_{xy}/(\sigma_{xx}\sigma_{yy}-\sigma_{xy}\sigma_{yx})\approx
\sigma_{xy}/\sigma_{xx}\sigma_{yy}$, is only weakly
temperature dependent.
However, this is not the case in
HTSs,\cite{donbook} where typically the Hall resistivity
decreases
sharply with increasing temperature.
As discussed in the Introduction,  it
has been suggested (in the context of the RVB ground state) that
the
scattering processes which involve  $\rho_{xx}$ and $\rho_{xy}$
are
intrinsically different, and that the cotangent of the Hall
angle,
$\cot\Theta_H\equiv\rho_{xx}/\rho_{xy}$ is universal and
proportional
to $T^2$ (Ref.\ \onlinecite{pwa-hall}).
Much of the experimentally available data focus
on the Hall coefficient,
$R_H=\rho_{xy}/B$, where $B$ is the applied magnetic field,
although
the above proposal has led a number of authors to plot
$\cot\Theta_H$. However,
we find that an examination of the behavior of $\sigma_{xy}(T)$,
a
quantity more  directly related to theoretical calculations,
provides
more insight into the underlying physical origin of the measured
anomalous Hall effect behavior.

Since $\rho_{xx}$ typically varies linearly with $T$ for
$T>T_0$,
$\sigma_{xy}$ must decrease faster than $T^{-2}$ in order
to bring about the temperature dependent behavior of $R_H$.
Examination of
the experimental data presented in Figs.\ \ref{fig:sigmaxyt3ito}
and
\ref{fig:sigmaxyt3lsco} shows that this is indeed the case at
high
temperatures, where $\sigma_{xy}$ is seen to vary as $T^{-3}$
for both
the 123 and 214 systems. However, experiment also shows
deviations
from this high temperature behavior in many systems. From Eq.\
(\ref{eq:sigmaxyemp}) it is clear that if $T_0$ is temperature
independent,
$\sigma_{xy}$ should indeed vary as $T^{-3}$. However, in PS
regime this
is not the case, since $T_0\sim 1/T$ and
hence the result in Fig.\ \ref{fig:sigmaxyt3ito} is
to some extent in contradiction with our result. However, the
inset of
the Figure shows clearly that in 
YBa$_2$Cu$_3$O$_{6.63}$ 
$\sigma_{xy}$ obeys $T^{-4}$ quite well up to temperatures $T\sim 200K$,
where the first
deviations from $T^{-4}$ behavior occur. In fact,
only well above this temperature does $\sigma_{xy}$ become
proportional to $T^{-3}$,
in agreement with our predictions.
One can observe similar deviations at other doping levels in
this system.
Note also that $\sigma_{xy}\sim T^{-3}$ is
obtained also well above $2\pi T_0$, for $T_0\sim 1/T$ (see Table 
\ref{tab:coefs}).
We draw two important conclusions. First, $\sigma_{xy}$ has a
crossover
from $T^{-4}$ to $T^{-3}$ behavior, the details of which depend not
only on whether one
is above $T_0$, but also on the temperature regimes considered.
Second, and
more importantly, the ``disagreement'' in the experimental data
in Fig.\
\ref{fig:sigmaxyt3ito} and its inset show clearly that one
should not
use simple minded power law fits, as is often done in the
literature, but
more complex formulae like Eq.\ (\ref{eq:sigmaxyemp}), which
include
all necessary crossovers. In addition, we note that one must be
very careful
when fitting to the Hall effect data, since impurities play an
important role,
and their presence can lead to large positive intercepts of
$1/\sigma_{xy}$
as a function of $T^3$, such as that seen in Fig.\
\ref{fig:sigmaxyt3lsco}.
Moreover, as mentioned above,
the chain contribution in various materials is non-trivial and may
obscure
a crossover in the in-plane Hall effect. Finally, we remind the
reader
that Eq.\ (\ref{eq:sigmaxyemp}) is obtained in the limit
$\omega_{sf}\ll T$.
This limit is not well satisfied in highly overdoped materials
where a somewhat
different sequence of crossovers may occur.

Further
insight into the role played by anisotropic quasiparticle
scattering
may be obtained using a phenomenological
model for the variation in the mean
free path as one goes around the FS.
For simplicity, we consider  a
representative FS and anisotropic quasiparticle behavior which
exhibits
a four-fold symmetry. We assume a
cylindrical FS with approximately constant $k_f$, but with an
anisotropic effective mass $m_k$.
We parametrize the mean free path (MFP) by assuming that there
are hot
regions on the FS at 90$^o$ with respect to each other, and
choose as a
representative FS that depicted in Fig.\ \ref{fig:fs}. In the
extended
zone scheme this FS is approximately a large distorted circle,
centered around $(\pi,\pi)$, with hot spots near ${\bf
k}=\pm(\pi,0)$
and ${\bf k}=\pm(0,\pi)$.
Then the MFP around the FS is given by:
\begin{equation}
{1\over \ell(\theta)} = {1\over \ell_{hot}}
\left[{1+a\cos(4\theta)\over 1+a}\right]
\label{eq:param}
\end{equation}
where the anisotropy parameter, $a$, varies between 0 and 1
($a\approx 1$ for a highly anisotropic scattering rate
and $a\approx 0$ for weak anisotropy),
and $\ell_{hot}$ is the MFP at a hot spot
on the FS. Note that, in general, both
$a$ and $\ell_{hot}$ are functions of temperature. For
simplicity we shall assume $a=(1-r)/(1+r)$, where
$r=\ell_{hot}/\ell_{cold}\equiv \ell(0)/\ell(\pi/4)$.
This particular choice may not be the most suitable for
an effective interaction which is sharply peaked  at ${\bf Q}$,
since
for long correlation lengths, as one moves away from a hot
spot $1/\ell$ decays faster with $\theta$ than  is assumed here.
However, if the effective interaction
(\ref{eq:mmp}) has incommensurate peaks in momentum space, then
the anomalously scattered (hot) regions
are large, $1/\ell$ is a slower function of $\theta$, and Eq.\
(\ref{eq:param}) is quite reasonable approximation.
Therefore, one should consider $\ell_{hot}$ and $\ell_{cold}$ here in 
a broader sense, recognizing that under certain circumstances it may 
not be possible to identify them with the 
results quoted in the previous section.

Starting from the expressions, Eqs.\ (\ref{eq:sigmaxx0}) and
(\ref{eq:sigmaxy0}),  for $\sigma_{xx}$ and $\sigma_{xy}$,
at $T\ll t$ one can perform a change of variables
and integrate over $\epsilon$.  The result for $\sigma_{xx}$ is:
\begin{equation}
\sigma_{xx}={e^2k_f\over 2 \pi^2} \int d\theta
\,\ell\,\cos^2\theta
\end{equation}
where $\ell=\tau(\theta) k_f/m(\theta)$
and $\theta$ is the angle between the electrical field and ${\bf
k}(\theta)$. On using the parameterization (\ref{eq:param}), we
find:
\begin{equation}
\sigma_{xx}={n e^2 \tau_{hot}\over m_{hot}\sqrt{r}}
={n e^2 \tau_{cold}\sqrt{r}\over m_{cold}}
\label{eq:sigmaxx}
\end{equation}
where $n=k_f^2/2\pi$.
In similar fashion the Hall conductivity is found to be:
\begin{equation}
\sigma_{xy}=\left( {e^3 B \over 2 \pi^2} \right) \int d\theta
\ell\cos\theta \left({d\over d\theta} \ell\sin\theta\right)
\end{equation}
which leads to:
\begin{equation}
\sigma_{xy}=\left({e^3B\over 2\pi}\right){\ell_{hot}^2}
{1+r\over
2r^{3/2}}
=\sigma_{xx} eB\ell_{hot} {1+r\over 2 r}
\label{eq:sigmaxy}.
\end{equation}
Eqs.\ (\ref{eq:sigmaxx}) and (\ref{eq:sigmaxy}) may be combined
to
yield simple expressions for the Hall coefficient and Hall
angle:
\begin{equation}
R_H\approx {\sigma_{xy}\over \sigma_{xx}^2 B}=
{1\over en} {1+r\over 2\sqrt{r}}
\label{eq:hallcoef}
\end{equation}
and
\begin{equation}
\cot\theta_H={\sigma_{xx}\over \sigma_{xy}}={m_{hot}\over
eB\tau_{hot}}\, {2r\over 1+r}\label{eq:pheno-cot}
\end{equation}
Note that the above expressions depend only on the two
temperature
dependent
parameters $r$ and $\ell_{hot}$. These two quantities can be
directly
probed by measuring the Hall coefficient (\ref{eq:hallcoef}) and
the resistivity $\rho_{xx}\approx \sigma_{xx}^{-1}$, since $R_H$
depends only on $r$ and the product $\rho_{xx}R_H$ yields
$\ell_{hot}$.

In the case of high anisotropy, i.e.,
$r\ll 1$, the above quantities take an especially simple form:
\begin{eqnarray}
\sigma_{xx}&=&{n e^2\sqrt{\tau_{cold}\tau_{hot}\over m_{hot}
m_{cold}}}
\label{eq:sigxxsmallr}\\
\sigma_{xy}&=&\sigma_{xx} {\omega_c \tau_{cold}\over 
2}\label{eq:sigxysmallr}\\
R_H&=&{1\over 2 n e}\sqrt{\tau_{cold}/m_{cold}\over
\tau_{hot}/m_{hot}}\label{eq:rhoh}\\
\cot\theta_H&=&{2\over \omega_c \tau_{cold}}\label{eq:tcoeftau}
\end{eqnarray}
where $\omega_c$ is the {\em cold} cyclotron frequency,
$\omega_c=eB/m_{cold}$. Our result, Eq.\ (\ref{eq:tcoeftau}) for
the
Hall angle, provides a natural explanation of why the
superconducting cuprates with a variety of different behaviors
for
$\rho_{xx}(T)$, often display a quite similar behavior of
$\cot\Theta_H$.
Note that the Eq.\ (\ref{eq:sigxxsmallr}) is exact for any $r$.
It is clearly consistent with our conclusion in Sec.\ \ref{sec:theory} that 
the linear in $T$ resistivity in NAFLs is due to a fine balance between 
cold and hot regions. In addition, with the help of Eq.\ 
(\ref{eq:lamecoldlargek}) one easily verifies that Eq.\ 
(\ref{eq:sigxysmallr}) is consistent with Eqs.\ (\ref{eq:rhoxx})
and (\ref{eq:sigmaxyemp}).

One can use Eqs.\ 
(\ref{eq:sigmaxx})-(\ref{eq:pheno-cot}) to
deduce the temperature dependence of $(m/\tau)_{hot}$ and
$(m/\tau)_{cold}$ from the experimentally measured $\rho_{xx}$
and $\rho_{xy}$. Only one undetermined parameter enters into
this procedure, namely $n$, which in some cases can be obtained 
experimentally from the high temperature limit of $R_H$, since in this 
case $r\rightarrow 1$, and the Hall coefficient is temperature 
independent.
The quantity. $R_\infty\equiv R_H(T\rightarrow\infty)$ 
is well defined for many overdoped
materials,\cite{batlogg} but has been determined for only a few
underdoped
materials.  Whenever experimental data are not available
we make the best estimate for $R_\infty$:
for example,
in the 214 family we extrapolate $R_H(T)$ to $T=T_{cr}$
and for the 124 material we assume $R_\infty$ comparable to the
result
found for

YBa$_2$Cu$_3$O$_{6.63}$ (Ref.\ \onlinecite{ong-500}),
which has comparable magnetic
properties. However it
is important to stress that as long as one has the correct order
of
magnitude for $R_\infty$, qualitatively the results are
virtually
the same.


We proceed to extract quasiparticle scattering rates by fitting
Eqs.\
(\ref{eq:sigmaxx})-(\ref{eq:hallcoef}) to the transport
measurements
considered above. 
Fig.\ \ref{fig:668rates} shows the
scattering rates for YBa$_2$CuO$_{6.63}$
in the cold and hot regions of the FS
(in the same dimensionless units) 
as a function of temperature.
Clearly in the hot region the scattering rate is much larger;
it is approximately linear in $T$ for $T\leq T^r_*\sim 200$K,
and becomes weakly $T$ dependent at higher temperatures.
As noted above, for this system $T_*^r\approx T_\star$.
At the same time $(m/\tau)_{cold}$ barely reflects the crossover
at $T_*^r$ as may be seen in Fig.\ \ref{fig:668cot}, although clearly
both scattering rates are affected as the system approaches the
PG regime. Note that these scattering rates closely
resemble those shown in Fig.\ \ref{fig:ratesPS}. In particular, the
crossover to a constant scattering rate of $(m/\tau)_{hot}$ is strikingly
similar to the calculated one for $\Delta k\approx 0$, shown in Fig.\
\ref{fig:ratesPS}.

The results for the 124 compound are given in Fig.\
\ref{fig:1248rates}, where we have taken $\rho_{aa}$, which
does not include chain contribution and exhibits
an in-plane resistivity resembling that in underdoped
materials,
and used the measured values of $R_H$. The results are
similar to those found for  
YBa$_2$Cu$_3$O$_{6.68}$, although
in the measured temperature regime ($T<300$K) the scattering
rate at the hot spot does not quite saturate.
However, we find that, as $T$ increases,
$(m/\tau)_{cold}$ becomes approximately proportional
to $T^2$ in this material; on extrapolating this behavior of
$(m/\tau)_{cold}$ and the experimentally obtained resistivity,
we find the same behavior of $(m/\tau)_{hot}$, albeit with a 
somewhat higher value of $T_*^r\approx 250$K.

Finally in Fig.\ \ref{fig:123o7rates} we show the scattering
rates
in the cold and hot regions for YBa$_2$Cu$_3$O$_7$:
 here the scattering rate in the hot region is once more seen to 
saturate, but at
a much lower temperature corresponding to 
$T_*^r\approx 120-130$K. Both this value
and that found in 124 and YBa$_2$Cu$_3$O$_{6.68}$ are very close to $T_\star$,
found by Barzykin and Pines for these materials (see Table
\ref{tab:crossovers}).
Although it is not easy to verify directly in NMR experiments, because 
both $T_{cr}$ and $T_*$ are close to $T_c$
for YBa$_2$Cu$_3$O$_7$, this result provides a further indication of the
presence of the PG phase even in optimally doped 123 materials.

A similar analysis can be performed in single CuO layer
materials:
two panels of Fig.\ \ref{fig:214rates}
show the scattering rates at doping levels $x=0.10$, $x=0.15$ in
214
compound. Much like YBa$_2$Cu$_3$O$_{6.68}$, one observes the saturation
of the
scattering rate in the hot region and it is only for a
particular
anisotropy level that the resistivity in the optimally doped 214
material is approximately linear in $T$ below $T_*^r$. In fact,
as
pointed out in Ref.\ \onlinecite{batlogg}, a crossover in
resistivity exists even in this material, but 
is so minor that one usually assumes
that $\rho_{xx}\propto T$ at all temperatures.
As mentioned before, in
the 214 family  $T_*^r$ is considerably  larger than $T_{*}$.

Finally we consider the overdoped single layer materials:
Fig.\ \ref{fig:tlrates} shows the scattering rates for the
overdoped
15K Tl 2201 compound. First, one
immediately notices that the anisotropy of scattering is not nearly
as
pronounced as in previous graphs. Note that in this case 
$R_\infty$ is very well known and thus one can
consider this reduced  anisotropy in scattering rates
quantitatively as well. We
remark that both scattering rates have nearly as strong $T$
dependence as in the previous cases, over the entire  $T$ range.
Moreover, the scattering rates shown here again closely resemble
the calculated result for larger $\Delta k$ in Fig.\
\ref{fig:ratesMF}.

\section{Numerical Results}
\label{sec:numerics}
\typeout{numerics on page \thepage}

In the previous sections we have established the relationship
between
the crossovers in the magnetic spectrum of cuprates and their
transport
properties analytically. However, it is unclear whether
such an analysis, based on
back-of-the-envelope calculations,
can account for the experimental results quantitatively.
In this section we present our numerical results for the
transport
coefficients in NAFLs: we show that for
spin-fluctuation
and band parameters  chosen from NMR, INS and ARPES
measurements,
the NAFL model indeed reproduces
experimentally obtained results. 
We  compare our results to experiments performed
on several compounds of interest, examine the sensitivity of our
results to the input parameters of the theory, and
address several important questions regarding realistic
materials,
including the strong a-b plane anisotropy observed in a number
of
bilayer compounds.

The method used to obtain the results presented here has already
been
discussed elsewhere\cite{bps}
and hence we only review it briefly. We study the transport
in NAFLs by solving the Boltzmann equation (BE) numerically. We
obtain the displacement of the quasiparticle (fermionic)
distribution function,
\begin{equation}
f_0(k)-f(k)=\Phi_k\left({\partial f_0(k)\over \partial
\epsilon}\right)
\end{equation}
as a function of momentum $k$, temperature and frequency on a
fine mesh
of points in the Brillouin zone, using a standard relaxation
method.
The transport coefficients are then obtained from
\begin{equation}
{\bf j} = e\sum_k \Phi_k {\bf v}_k \left({\partial f_0\over
\partial \epsilon}\right)
\label{eq:sigmaij}
\end{equation}
where ${\bf v}=\nabla \epsilon_k$. We assume that the
interaction present in the collision integral in the BE, is
given
by  Eq.\ (\ref{eq:mmp}).
The numerical method can be described as follows:
we start from a reasonable choice
of $\Phi_k$, which we then use to calculate the collision
integral,
which  in turn provides for a new value of $\Phi_k$.
The iteration procedure
is stopped when the difference between two
subsequent values of $\Phi_k$
is smaller than a given tolerance.

In our earlier numerical work,\cite{SP,bps} we demonstrated the
viability of a NAFL description of transport in the cuprates.  By using
spin-fluctuation spectra seen in NMR experiments, we were able to obtain both
qualitative and quantitative agreement with the analysis presented in
Sec.\ \ref{sec:theory} and Sec.\ \ref{sec:experiment}.  However, for the band
parameters, doping level, and spin-fluctuation spectrum we assumed for
optimally-doped YBa$_2$Cu$_3$O$_7$, the crossover temperature, $T_o$, was high
compared to that seen experimentally, while $\rho_{xx}$ exhibited FL behavior up
to temperatures
$\sim 200$K.  Because $T_o$ was large, on extrapolating our linear in T results
for $\rho_{xx}$ to zero temperature, we found a large negative intercept.
Moreover, although the calculated Hall coefficient decreased with increasing
temperature, it displayed a far weaker temperature dependence than that seen
experimentally.  Our results thus resembled those found for overdoped cuprates,
rather than an optimally doped sample.

On the basis of the analysis presented in \ref{sec:theory}, we can identify
several potential causes for these discrepancies.  First, the actual value of
$\omega_0 \equiv \omega_{\rm SF}\xi^2$ in optimally doped 123 is $\alt
60$meV,\cite{zbp} rather than the larger value, $\omega_0 = 76$meV, adopted
in these calculations.  Second, the spin fluctuation spectrum was assumed to
possess a commensurate peak, rather than the four incommensurate peaks which a
recent analysis of NMR and INS results suggests might provide a better fit to
experiments on the   system.\cite{zbp}  Third, the doping level was taken to
be $n \sim 0.25$\%, while the LDA calculations of bonding and antibonding in
bilayer materials discussed below suggest a considerably lower doping level
might be more appropriate.  Each of these effects acts to bring about either a
lower value of $T_o$, an increase in the linear in T regime of $\rho_{xx}$,
and/or an increased temperature dependence of $\rm R_H$.  To explore their
combined influence, we carry out a numerical solution of the BE for a
representative ``standard'' 123 material for which $n = 20$\%, $\omega_0 =
60$meV, and $\omega_{\rm SF}(K) = 60 + 0.6 T(K)$; 
the band spectrum was assumed
to be $(\rm t = 0.2$eV; $\rm t' = -0.35t)$, which is slightly different 
from the unrenormalized LDA spectrum,\cite{OKA}
while the incommensurate peaks in the spin spectrum are 
assumed to lie along the
diagonals, at ${\bf Q}_i = {\bf Q} + (\pm 0.1\pi, \pm 0.1\pi)$, as proposed by
Zha et al.\cite{zbp}

In Fig.\ \ref{fig:rxx123} we compare our calculated
resistivity as a function  of temperature with the
experimental results obtained for optimally-doped YBa$_2$Cu$_3$O$_7$. The
coupling  constant $g$, which sets the scale for the magnitude of the 
resistivity, has been adjusted to $g=0.48$eV, in order to
obtain the best fit.  Clearly, the calculated
result agrees very well with the data down to temperatures only slightly
above
$T_c$. Note
that the extrapolated value of $\rho_{xx}$ at $T=0$ is now quite
small, as is seen in the experiment. On the other hand, the Hall coefficient,
shown
in the inset of the figure, is strongly temperature dependent, with a relative
decrease which is approximately the same as that seen experimentally
in the  temperature interval, 150-400K,
although the exact shape of $R_H$ is not the same.
However, as mentioned
above, a better knowledge of the band structure, including the
chain contribution,
and of the effective interaction $V_{eff}$ for ${\bf q}$ away from
${\bf Q}_i$  might well remove this
problem.  Moreover, a quick look at Eq.\ (\ref{eq:rhoh}), as well as 
Eqs.\ (\ref{eq:rhoxx}) and (\ref{eq:sigmaxyemp}) indicate that the FS 
reduction (i.e., decreasing $\Delta k_{max}$)
with the approaching PG phase at low temperatures, should 
provide for an additional temperature dependence of $R_H$.
Our calculated $R_H$ saturates at higher $T$, to
a somewhat higher value than that seen experimentally, implying that a 
different value of the doping level should be used; nevertheless, our
choice of parameters would seem quite 
reasonable (see Eq.\ (\ref{eq:hallcoef}).
Another reason for a discrepancy in the shape of $R_H$ lies in
the fact that we have calculated this
quantity assuming a tetragonal lattice,
rather than the orthorhombic one found in YBa$_2$Cu$_3$O$_7$.
Recent measurements\cite{anton} on twinned
 and detwinned crystals of this
material show a considerable difference in
$R_H$, indicating the importance of including
the a-b plane anisotropy in the calculation.
We return to this point at the end of this section.

The doping level used in this calculation is somewhat low.
If one assumes that every  Oxygen atom doped into the 123 system
extracts
a single hole from CuO planes, the doping level of the optimally
doped
123O$_{6.93}$ should be somewhat higher, $n\approx 23\%$.
On the other hand, simple arguments, based on LDA
calculations, suggest that in bilayer materials, such as 123
compounds,
there are two bands, bonding and antibonding, separated by
twice the hopping matrix element between the CuO layers,
$t_\perp$ (see Refs.\ \onlinecite{OKA,girsh}).
If one assumes  that $t_\perp$ is weakly momentum dependent, one finds
the top (antibonding) band is more heavily doped, while the
bottom (bonding) band should be close to half filling,
regardless of the Oxygen content.\cite{girsh}
Since the interaction between quasiparticles
is much weaker in the metallic state of cuprates
than it is in the insulating state, the nearly
half filled bonding band has a larger value of $v_f$ and should
provide the dominant contribution to the conductivity
in the bilayers (see Eq.\ (\ref{eq:rhoxx})).
Hence the doping level
chosen here, $n=20\%$, may be quite reasonable for 123 and 124
compounds.

To study the effect of doping on the
resistivity at fixed spin fluctuation spectra 
 we have calculated  $\rho_{xx}(T)$
for the same values of
the spin and band parameters as in our ``standard model,'' but
with the chemical potential adjusted to yield an
$n=30$\% doping level, and  compared it to the above case
in Fig.\ \ref{fig:incom}. 
Clearly, $\rho_{xx}$ ceases to be linear in
$T$ at lower temperatures, while continuing to exhibit typical FL behavior
as
$T\rightarrow
0$.
In the same figure we also show the influence of
incommensuration, by
plotting  $\rho_{xx}(T)$ for the same representative
set of parameters as above, but with a commensurate
spin-fluctuation
spectrum. Although the commensurate result has a
slope at higher temperatures which is remarkably similar to the
incommensurate
(representative)
 one (due to the fact that $\rho_{xx}$ is independent of
 $\omega_0$ above $T_0$ (see Eq.\ (\ref{eq:rhoxx})),
 we  again  observe FL-like
 behavior  at  lower  temperatures.
 In both cases the explanation is simple: enlarging $\Delta k_{max}$,
 leads to an increase in $T_0$; this can be achieved
 either  by increasing the doping
level, which increases the size of the FS, 
or by removing the extra hot spots brought about by
incommensuration. Fig.\ \ref{fig:incom} demonstrates
quite clearly that the behavior of
the resistivity is sensitive to the choice of band
parameters and leads us to conclude that 
 self-consistent calculations which take into account changes in the 
quasiparticle interaction should be used
in comparing with experiment.\cite{MP}
On the  other hand, assuming that the magnetic (band) parameters
are
well known, our model can be used to put constraints on the
allowed band
(magnetic) parameter values.

Next we examine the dependence of $\rho_{xx}$ and $\sigma_{xy}$ on
$\omega_{sf}=A+BT$. In Fig.\ \ref{fig:rxxwsf} we
plot the resistivity as a function of temperature for four
different values of $A$ and $B$: we show
the cases where $A=20$K and $B=0.2$,
$A=0.6$ and $B$ is the same,
the case where $B$ is 0.6
and $A$ is the same, and the
case where $\omega_{sf}=60+0.6T$. The band parameters are set to 
$t=250$meV, $t'=-0.45 t$ and the doping level is $n=15$\%; the coupling 
constant is set to $g=1$eV.
In agreement with  our analysis in Sec.\ \ref{sec:theory},
a large value of $A$,
corresponding to an enhanced damping of spin-fluctuations
due to spin-fermion
interactions, shifts the resistivity down; it
leads to a higher value of $T_0$ in Eq.\
(\ref{eq:rhoxx}) and only marginally
anomalous scattering rates even in hot
regions of the FS, in agreement
with the result presented in Fig.\
\ref{fig:tlrates}

On the contrary, increasing the value of $B$ only changes the
slope of the resistivity curves shown in Fig.\ \ref{fig:rxxwsf},
but does not much affect the behavior with respect to the origin.
  This is to be expected, since according to Eq.\ (\ref{eq:q_k})
the term $BT$ in $\omega_{sf}$ acts in conjunction with the
$\pi T$ term in $\sqrt{\pi T + \omega_{sf}(1+(\Delta
k)^2\xi^2)}$ to bring about linear in $T$ resistivity. However,
since in all experimentally studied samples $B\ll \pi$, the
effective value of $T_0\sim \omega_0 (\Delta k_{max})^2/2(\pi +
B)$ differs only slightly from that quoted in Eq.\
(\ref{eq:t0}). As noted in Sec.\ \ref{sec:theory}, for practically all
cuprates
$BT$ is never large enough to produce the $T^{1/2}$ behavior of the
resistivity discussed there, which is found at $T\sim \omega_0
(\Delta k_{max})^2$, provided the FS is large enough. Hence, the crossover at
$T^r_*$,  where a curvature resembling $T^{1/2}$ is found in many
underdoped samples at $T\geq T_*$, has to be attributed to 
a small value of
$\Delta k_{max}$ resulting from the FS evolution. Finally we note that
the Hall  conductivities for all of the above cases obey a
$T^{-3}$ law quite well, as shown in the Inset of the figure,
although
due to the large value of $T_0$ the Hall resistivity may be only
weakly temperature dependent.

We turn now to the $z=1$ PS regime. As indicated in Sec.\
\ref{sec:theory}, the resistivity is linear in $T$ for $T>T_0$
regardless of whether or not $T_0$ is temperature dependent. However,
since $\xi\sim 1/T$ in this regime, the condition $\xi\Delta k
<1$ is satisfied over most of the FS at relatively low
temperatures.
In this case the resistivity due to spin-fluctuation
scattering
tends to a constant, since  $T\gg 2\pi T_0$.
To further explore the PS regime, we solve
the B.E. numerically for a representative
member of the 2-1-4 system, [$\hat{c}=\omega_{sf}\xi=50$meV,
$\xi^{-1}=0.1 + T/1000$K, $g=1$eV,
$n=15$\%, $t=250$meV] and
several values of $t^\prime$ ($-t^\prime/t=0.2$, 0.3 and
0.4 respectively). Our results are shown in Fig.\
\ref{fig:rxx214}a.  In all of the cases displayed there we 
have assumed that the
effective interaction has incommensurate peaks at ${\bf Q}_i={\bf
Q} + (\pm \Delta Q,0)$ and ${\bf Q} + (0,\pm \Delta Q)$, with $\Delta 
Q=0.25\pi$, in order to be
consistent with neutron scattering experiments performed in 214
materials at sufficiently high doping levels.\cite{zbp}
For $t^\prime=-0.4t$ the FS is relatively large,
as is the corresponding $T_0(T)$. As a result one
observes a crossover to linear in $T$ resistivity at $T\sim
200$K
from the low $T$ Fermi liquid like behavior.
For $t^\prime=-0.2t$ the FS
is always close to the Brillouin zone boundary and hence the
effective $\Delta k_{max}$ is small, $\Delta k_{max}\sim 0.1$.
Hence the condition $T=2\pi T_0(T)$ is satisfied for $T\sim
1000$K, and the resistivity deviates from linearity already at
$T=150$. Obviously, in the intermediate case $\rho_{xx}\propto T$
over an extended temperature range. We conclude that the
resistivity is very sensitive to changes in band parameters,
especially when $T_0$ is a strong function of temperature.

The parameters used in  Fig.\ \ref{fig:rxx214}a are close
to those found for
La$_{1.85}$Sr$_{0.15}$CuO$_4$ material in the PS 
regime.\cite{BP,zbp} In Fig.\ \ref{fig:rxx214}b we compare
$\rho_{xx}(T)$, obtained  for $t'=-0.33t$ (solid line), to experiment.
The coupling constant has been adjusted to yield the same slope
as the experimental curve.
Obviously even in this case the calculated resistivity exhibits
the features shown in Fig.\ \ref{fig:rxx214}a, but the fit is quite
good, especially keeping in mind the fact that the experimental
result is subject to disorder.
On the other hand, when one compares $\cot\Theta_H$ in these two
cases (solid line in Fig.\ \ref{fig:214cot}), 
the fit, although remaining reasonable, is not as good.  The
curvature of the calculated result (and lack thereof in the
experimental result)
implies that the cold regions
of the FS are too small in the calculation, whence
 $\cot\Theta_H\sim T^{1.6}$, rather than the $T^2$, behavior frequently
seen, although some samples of this material have the smaller
exponent in $\cot\Theta_H$ found here. 
However, even for a small value of $t^\prime$,
the addition of a Fermi liquid
term, $\chi_{FL}=\chi_0(T)/(1-i\pi\omega/\Gamma(T))$, 
to the effective interaction, makes the
resistivity appear linear in $T$ up to 
higher temperatures and the 
agreement with experiment for the Hall effect much better,
as shown by the dashed line in Figs.\ \ref{fig:rxx214} and 
\ref{fig:214cot}. In both cases we have used $\Gamma(T)=(0.5 - T/2500)$eV 
and $\chi_0(T)=1 + 0.4 T/100$states/eV, the values obtained from the 
fits to NMR measurements.\cite{monien} 
This is as expected, since
adding a Fermi liquid term affects predominantly the cold
regions
of the FS and its effect is similar to that of the strong band
renormalization present in the PS regime.

We next consider the issue of the a-b plane anisotropy. Although found in
practically all cuprates, this anisotropy is especially pronounced in materials
with CuO chains, leading to quite different transport results along different
crystallographic directions. In principle, these are two reasons for an
anisotropic resistivity: the chains can form a conducting band, or the chain
bands hybridize with Oxygen bands in CuO plane, producing an anisotropic
quasiparticle effective mass.  
Since the chains in, e.g., YBa$_2$Cu$_3$O$_7$ are very close
to the CuO planes, and the Oxygen p-orbitals are rather large, we believe that
the latter effect must dominate. In order to simulate its effect on the in-plane
resistivity we consider a 2D band with anisotropic hopping integrals: we alter
the first term in Eq.\ (\ref{eq:dispersion}) to read
\begin{equation} -2t(\cos k_x + a \cos k_y)
\label{eq:aniso}
\end{equation} In Eq.\ (\ref{eq:aniso}) the anisotropy parameter $a$, lies
between 0 and 1; obviously, $a=1$ in
tetragonal systems. A FS for an anisotropic system with $a=0.55$ and the doping
level $n=10$\% is shown in Fig.\  \ref{fig:fs1}. Although this shape of the FS
is somewhat incompatible  with that observed in, e.g., the 
124 material, one must
keep in mind that strictly speaking a single band model, such as that
considered here, is an oversimplification; we only
use the present model in order to qualitatively explain the physics in the above
material.

In Fig.\ \ref{fig:aniso} we show the resistivities for the
current running along x and y directions for the
band depicted by the FS in Fig.\ \ref{fig:fs1}
We assume that $\omega_{sf}(K)=20+0.2T$,
which is close to $\omega_{sf}$ found in 124 materials. However, we 
retain the $z=2$ scaling regime in order to make the comparison with 
previous results easier. Clearly, not only are the magnitudes
of the two resistivities different, but their temperature dependence
is also different;
indeed our results for $\rho_{xx}$ and $\rho_{yy}$ resemble
those found in the 124 material.
The lower curve ($\rho_{xx}(T)$)
appears to have a considerable negative intercept, indicating a
large value of
$T_0$, as in overdoped materials. The upper curve,
$\rho_{yy}(T)$, exhibits a crossover at $T \sim 300$K, 
similar to that observed experimentally at
$T^r_*$ in underdoped  cuprates.
 This quite different behavior arises from  the
modified
positions of the hot spots: as may be seen on examining the
FS shown in fig.\ \ref{fig:fs1}, the hot spots are now located
at points $(1,\pm \pi)$ and  $(\pi-1,0)$ and symmetry related
points on the FS, i.e., they are asymmetric with respect to the
inversion of $x$ and $y$ axis
and cannot be obtained by drawing
the usual magnetic Brillouin zone boundary.
For current running along the $x$ direction,
a large segment of the FS, roughly along the $(1,y)$ direction,
can be displaced in transport (the vertex function $v_x$ is
large at
these $k$ vectors) and more importantly, these segments of the
FS are far from the hot spot at $(1,\pm \pi)$, so that it is
necessary to balance
the hot and cold regions,
to arrive at a $\rho_{xx}$ which is linear in $T$. 
For current running along the $y$ direction, the vertex
$v_y$ is small almost everywhere on the FS (see Eq.\ (\ref{eq:sigmaij})
 and the hot spots sit in the
 middle  of the region of the FS where $v_y$ is the largest, further 
reducing the conductivity in this case.
Effectively the  FS appears to be small, producing
 the    anomalous (non-linear with negative curvature)
$\rho_{yy}(T)$
 seen in the figure.   Note that
 the Hall conductivity is proportional to $T^{-3}$, as seen in
Fig.\ \ref{fig:hallaniso}. We also note that our result for
$\sigma_{xy}$
is the same to within a few percent for both current directions, which
serves as a nice check of the quality of numerics in this case.

We make another important comment regarding this result: unlike
the
case of 124 material, where $T_*^r$ is very close to $T_*$,
so
that the anomalous behavior of the resistivity occurs at least
in the
pseudoscaling ($z=1$) regime if not in the pseudogap one, here
we have
similar behavior with $z=2$. We again conclude that the
crossover at
$T^r_*$ is closely related to the effective
size of the FS and is less closely related
to a particular magnetic regime.\cite{c-p-s}
However, we must keep in mind that
the FS here is reduced artificially (by introducing a very large
anisotropy). In reality, magnetic
fluctuations alter the shape of the FS as the system enters
into the pseudogap regime. Moreover, one cannot obtain a similar
anomalous behavior of $\rho_{yy}$ using a large Fermi surface
and the
same spin-fluctuation parameters as those utilized here; and hence
this result serves as a further indication
that the pseudogap regime involves an evolution of the FS.

\section{Conclusions}
\label{sec:conclusions}
\typeout{conclusion on page \thepage}

We have performed an analysis of the in-plane longitudinal and
Hall conductivities (resistivities), in terms of the NAFL model, 
using both 
perturbation theory and numerical methods. From our perturbation
theory results we obtained an analytical  expression
for the scattering rate as a function of temperature for an
arbitrary point on the  FS, as well as empirical expressions for
 $\rho_{xx}$ and $\rho_{xy}$. The results are then used
to study analytically the crossovers in transport in various
 temperature regimes.
Assuming a particular geometry of the FS,  we
 obtained  phenomenological  expressions  for both $\rho_{xx}$
and  $\rho_{xy}$ which we used to analyze the experimental
results
and gain further insight into the scattering rates.  Finally, we
performed a set of numerical calculations, verifying 
for a realistic set of input parameters that 
NAFL model yields results in agreement with experiment.

Our analytical results display a relatively strong anisotropy of
scattering rates around the FS, leading to a complex morphology of both
$\rho_{xx}$ and $\rho_{xy}$, in contrast to the common belief
that in all cuprates $\rho_{xx}\propto T$ and $\rho_{xy}\propto
1/T$. A close examination of the available experimental
data shows that the
 {\em universality} of the temperature behavior of both
 $\rho_{xx}$ and $\rho_{xy}$
 is  quite   marginal over larger temperature regions,
  i.e., only extremely clean,
optimally doped samples of 123
show $\rho\sim \rho_0 + \alpha T$ with $\rho_0\sim 0$ and
$\cot\Theta_H\sim T^2$.  Even optimally
doped 214 compounds show considerable discrepancy from this
behavior in
that the intercept of $\cot\Theta_H$ at $T=0$
is non-negligible. All other doping levels show more
or less pronounced deviations from the above {\em optimal}
behavior at low temperatures, in agreement with the NAFL model.
We therefore conclude that the {\em optimally}
doped compounds are rather  unique and the temperature
variation of their transport coefficients is 
somewhat accidental.
The calculated results for the scattering rates are found to be
in detailed qualitative agreement with those coming from the
fits to the experimental data.
It is remarkable how systematically the scattering rates vary with
temperature and doping and how precisely they reflect the
magnetic properties in cuprates.

Finally our numerical results show reasonable agreement with 
experiments for a realistic set of spin-fluctuation and band parameters
in both $z=1$ and $z=2$ scaling regimes.
We also verified that improved quantitative agreement can be achieved 
through the inclusion of strong coupling effects, which bring about a
temperature dependent quasiparticle band structure. In addition, 
we showed that the NAFL model provides for a natural explanation for the 
unusual temperature dependence of the resistivities along different 
directions in 124 material. 
Both the numerical results and the available experimental data
provide yet another  indication that NAFL is a strong candidate
for the explanation of the anomalous transport in cuprates.

It is straightforward to extend the NAFL model to optical frequencies 
and to take into account the influence of impurities and we will 
report on these results in future publications.

\acknowledgements

We are indebted to P.\ W.\ Anderson,
A.\ Chubukov, V.\ Barzykin, B.\ Batlogg, G.\ Blumberg, G. Boebinger,
P.\ Coleman, J.\ Giapintzakis, D.\ Ginsberg,
R.\ Hlubina, A.\ Leggett, M.\ Lercher, N.\ P.\ Ong,
Q.\ Si, C.\ Slichter, A.\ Shegoliev,
R.\ Stern, D.\ Strong, T.\ Timusk and  S.\ Trugman for stimulating conversations on 
these and related topics.
We thank the National Center for
Supercomputing Applications for a grant of computer time.
This research is supported in part by NSF through
grants NSF-DMR 89-20538 (MRL at UIUC) and NSF-DMR 91-20000
(STCS).

\vfill\eject
\typeout{table on page \thepage}

\begin{table}
\caption{Crossovers temperatures and spin fluctuation parameters
for
representative cuprates}
\label{tab:crossovers}
\begin{tabular}{|l|cccr|}
\hline
 ~& ~$T_c$~ & ~$T_*$~ & ~$T_{cr}$(K)~ &~\\ \hline
~La$_{1.9}$Sr$_{0.1}$CuO$_4$~   & 33 & 50  & ~$\geq$500~&~\\
~La$_{1.85}$Sr$_{0.15}$CuO$_4$~ & 39 & 85  & 325&~\\
~La$_{1.8}$Sr$_{0.2}$CuO$_4$~   & 30 & 75  & 120&~\\
~La$_{1.76}$Sr$_{0.24}$CuO$_4$~ & 25 & 35  & 100&~\\
~YBa$_2$Cu$_3$O$_{6.63}$~       & 60 & 180 & ~$\geq$600~&~\\
~YBa$_2$Cu$_4$O$_{8}$~          & 80 & 210 & 475&~\\
~YBa$_2$Cu$_3$O$_{7}$~          & 90 & 110 & 150 &~\\
\hline
\end{tabular}
\end{table}

\begin{table}
\caption{The transport coefficients in various temperature
regimes.}
\label{tab:coefs}
\begin{tabular}{|l|ccccr|}
\hline
        ~& $\sigma_{xy}$ & $\rho_{xx}$ & $\rho_{xy}$ & $\cot
\Theta_H$&\\ \hline

$T< T_0$        & $T_0/T^4$    & $T^2/T_0$   & const       &
$T^2/T_0$
&\\
$T_0<T<2\pi T_0$& $T_0/T^3$     & $T$         & $T_0/T$     &
$T^2/T_0$&\\
$T>2\pi T_0$    & $1/T^2$  & $T$         & const       & $T$
&\\
\hline
\end{tabular}
\end{table}

\begin{figure}
\caption{A model of a Fermi surface in cuprates (solid line) and
the
magnetic Brillouin zone boundary (dashed line). The intercepts
of two
lines marks the regions of the FS near
$(\pi,0)$, which
can be connected by the wavevector ${\bf Q}_i$. These regions
are most strongly
scattered into each other (see Eq.\ (\ref{eq:mmp})).}
\label{fig:fs}
\end{figure}
\typeout{figures on page \thepage}

\begin{figure}
\caption{Scattering rates, in arbitrary units,
as a function of $T$ in the mean field $z=2$
regime, for several values
of $\Delta k$, the displacement of the wavevector $k$ from a hot
spot.
In both plots we have assumed $\omega_{sf}=0.2 T$,
$\omega_0=600$K. The
curves correspond to (top to bottom)
$\Delta k=0$, 0.1, 0.25, 0.5, 1 and 2. Panel (a) shows the case
of a quite {\em large} FS,
$\Delta k_{max}=2$, and panel (b) that of a typical small one, $\Delta
k_{max}=0.5$}
\label{fig:ratesMF}
\end{figure}

\begin{figure}
\caption{Scattering rates, in arbitrary units,
as a function of $T$ in the pseudoscaling
$z=1$ regime, for several values
of $\Delta k$, the displacement of the wavevector $k$ from a hot
spot.
In both plots we have assumed $\omega_{sf}=0.2 T$,
$\omega_0=1000$K$/T$ and the
curves correspond to (top to bottom)
$\Delta k=0$, 0.1, 0.25, 0.5, 1 and 2. Panel (a) shows the case
of a {\em large} FS,
$\Delta k_{max}=2$, and panel (b) that of a small one $\Delta
k_{max}=0.5$}
\label{fig:ratesPS}
\end{figure}

\begin{figure}
\caption{The FS as a function of coupling constant $g$ in a NAFL.
The figure shows only one quarter of the Brillouin zone.
The curves are obtained using the same formalism and the same values of 
spin fluctuation and band parameters as in Ref.\ 35. 
The dashed-dotted, solid and dashed lines correspond to increasing 
value of $g$. As shown in Ref.\ 29, these curves also correspond to 
high, intermediate and low temperatures respectively. 
Note that the effect of the
interaction is to move the FS towards the magnetic Brillouin 
zone boundary (here depicted by the dotted line). Hence the interaction 
effectively reduces $\Delta k_{max}$.}
\label{fig:fsevolution}
\end{figure}

\begin{figure}
\caption{The measured
resistivity as a function of temperature, showing the deviation
from linearity in $T$ in underdoped, overdoped and
optimally doped cuprates. Panel (a) shows the results from
Refs.\ 44, 45 and 46, 
obtained in (top to bottom)  15K Tl 2201,
YBa$_2$Cu$_3$O$_7$, and 
YBa$_2$Cu$_3$O$_{6.63}$ compounds. The dashed line is a
guide to the eye and the arrow marks the crossover from pseudoscaling 
to pseudogap behavior at
$T_*$ in the underdoped and optimally doped material. Panel (b) shows
La$_{2-x}$Sr$_{x}$CuO$_4$ at three doping levels (top to bottom)
$x=0.22$, 0.15 and 0.10 (Ref.\ 36). 
The quantity plotted is $(\rho_{xx}(T) - \rho_0)/\alpha T$, where
$\rho_0$ and $\alpha$ are obtained by fitting the high $T$,
linear part of the resistivity.}
\label{fig:rhotrho0}
\end{figure}

 \begin{figure}
 \caption{The resistivity $\rho_{aa}$, 
 for current running along the
$a$ crystallographic axis, as a function of $T$ 
 in YBa$_2$Cu$_4$O$_8$  
material (Ref.\ 37). 
 The quantity plotted is
$(\rho_{aa}(T) - \rho_{PS})/
\alpha_{PS} T$, where
$\rho_{PS}$ and $\alpha_{PS}$ are obtained by fitting
$\rho_{PS}+
\alpha_{PS} T$ in the PS temperature regime ($200$K$<T<500$K; see
Table I).
The arrows mark the crossovers at $T_*$ and $T_{cr}$. Note that
$\rho_{aa}$
retains its linearity in $T$ above $T_{cr}$ with only minor
change in
slope.}
 \label{fig:rhotstartcr}
 \end{figure}

\begin{figure}
\caption{The reduced resistivity, $(\rho_{aa}(T) - \rho_0)/\alpha T$,
in YBa$_2$Cu$_4$O$_8$ (Ref.\ 37) 
$\alpha$ and $\rho_0$ are determined as in Fig.\
\ref{fig:rhotrho0}.
The bottom (top) set of data shows $\rho_{aa}$ ($\rho_{bb}$),
where $\rho_{aa}$ ($\rho_{bb}$) is obtained
with current running along $a$ ($b$) crystallographic directions.
Inset: $\rho_{aa}$ and $\rho_{bb}$ in YBa$_2$Cu$_4$O$_8$ material. Note that at higher $T$ 
$\rho_{aa}\sim 3 \rho_{bb}$.}
\label{fig:bucherrho}
\end{figure}

\begin{figure}
\caption{$\rho_{xx}(T)$ in underdoped
cuprates at low temperatures. The three sets of data are
shifted by 1 for clarity and correspond to
(top to bottom) 
YBa$_2$Cu$_3$O$_{6.68}$, YBa$_2$Cu$_4$O$_8$ and
La$_{1.9}$Sr$_{0.1}$CuO$_4$
(Refs.\ 36, 37 and 46). 
The dashed lines are guides to the eye.
The quantity plotted is $\rho_{xx}(T) - \rho_{xx}(0))/
\beta T^2$, where
$\rho_{xx}(0)$ and $\beta$ are obtained by fitting $\rho_{xx} =
\rho_{xx}(0)
+ \beta T^2$ at low $T$.}
\label{fig:rhobelowtfl}
\end{figure}

\begin{figure}
\caption{The inverse Hall conductivity as
a function of temperature in 123 and
YBa$_2$Cu$_4$O$_8$ compounds. The data correspond to the measured results
obtained in
(top to bottom) YBa$_2$Cu$_4$O$_8$, 
YBa$_2$Cu$_3$O$_{6.63}$ and YBa$_2$Cu$_3$O$_7$ (Refs.\ 37, 45 and 46). 
Note that at sufficiently high temperature all curves satisfy
$T^{3}$ quite well. Inset: the result for 
YBa$_2$Cu$_3$O$_{6.63}$,
showing
that $\sigma_{xy}\propto T^{-4}$ at low $T$.}
\label{fig:sigmaxyt3ito}
\end{figure}

\begin{figure}
\caption{The inverse Hall conductivity 
for the 214 compound. The data are
taken from  Ref.\ 36 
and correspond to doping levels (top to
the bottom)   $x=0.10$, 0.15, and 0.22. The 
curves show large positive
intercepts, suggesting a considerable
amount of disorder and/or a smaller FS in this system.}
\label{fig:sigmaxyt3lsco}
\end{figure}

\begin{figure}
\caption{The {\em hot} (top) and {\em cold} (bottom) scattering
rates in YBa$_2$Cu$_3$O$_{6.68}$ in dimensionless units, 
obtained  by fitting the  results of Ito et al$^{46}$ for $\rho_{xx}$
and $\rho_{xy}$, shown in the previous figures, to the
phenomenological formulae (\ref{eq:sigmaxx}) and
(\ref{eq:hallcoef}) (see text). The rates closely
resemble the calculated scattering rates shown in
Fig.\ \ref{fig:ratesPS}. The arrow marks the crossover at
$T_*$. 
} \label{fig:668rates}
\end{figure}

\begin{figure}
\caption{The scattering rate in the cold regions of
YBa$_2$Cu$_3$O$_{6.68}$ obtained from experimental results of Ito et 
al.$^{46}$
Note
 that, to within an additive constant, this scattering rate is
 the same as that found in our perturbation theoretic calculations.
The arrow marks the crossover at $T_*$,
 discussed in the text.
 }    \label{fig:668cot}
\end{figure}

\begin{figure}
\caption{The hot (top) and cold (bottom) scattering rates for 
YBa$_2$Cu$_4$O$_8$ obtained from the experimental results of Bucher et 
al.$^{37}$ There scattering rates resemble closely those found in
YBa$_2$Cu$_3$O$_{6.68}$, as might be expected from the similarities in
their magnetic behavior. The arrow marks the crossover at $T_*$.
}  \label{fig:1248rates}
\end{figure}

\begin{figure}
\caption{The hot (top) and cold (bottom) scattering rates for 
YBa$_2$Cu$_3$O$_7$ obtained from the experimental results of Ginsberg 
et al.$^{45}$
Note that $1/\tau_{hot}$  shows a crossover from the $\Delta
k=0$ result in Fig.\ \ref{fig:ratesPS} to the same result in
Fig.\
\ref{fig:ratesMF},  which should occur near $T_{cr}$,
marked in the figure by an arrow.
$1/\tau_{hot}$ weakly increases with temperature
for $T\ll T_{cr}$ (see Fig.\ \ref{fig:ratesMF}). $T_*$ is the
crossover temperature equivalent to those shown in Figs.\
\ref{fig:668rates} and
\ref{fig:1248rates}. Note that $1/\tau_{cold}$ is approximately
quadratic in temperature, in agreement with our perturbation theory 
calculations. 
}  \label{fig:123o7rates}
\end{figure}

\begin{figure}
\caption{The hot (top) and cold (bottom) scattering rates for 
two 214 compounds obtained from the experimental results of Hwang et 
al.$^{36}$
The top (bottom)
panel shows the results for $x=0.10$ ($x=0.15$) doping level.
In the bottom panel the crossover temperature $T_*$ is clearly
observed in $1/\tau_{hot}$, as expected in the NAFL model.
}
\label{fig:214rates}
\end{figure}

\begin{figure}
\caption{The hot (top) and cold (bottom) scattering rates for a 
15K 2201 Tl compound obtained from the experimental results of 
Mackenzie et al.$^{44}$
Note that the anisotropy of scattering is considerably
reduced in overdoped materials, due to the smaller
correlation length and the strong fermionic damping,  manifested in 
a larger value of $\omega_{sf}$ (see text).}
\label{fig:tlrates}
\end{figure}

\begin{figure}
\caption{A comparison of our calculated 
result for $\rho_{xx}(T)$ with the experimental result of Carrington et 
al.$^{20}$
The coupling constant $g\approx 0.48$eV has been adjusted
to yield the best fit. Note that in this case $T_0$, Eq.\ (\ref{eq:t0}),
is of order 20K and hence one observes a very
small negative intercept of the
resistivity, $\sim 20 \mu\Omega$cm, as seen in experiment. Inset: the
Hall resistivity as a function of temperature in comparison with
experiment for the same system. }
\label{fig:rxx123} \end{figure}

\begin{figure}
\caption{The sensitivity of $\rho_{xx}(T)$ with respect to the band
and incommensuration parameters. The solid line shows the same
$\rho_{xx}(T)$ as in Fig.\ \ref{fig:rxx123} with $g=1$eV.
The dashed and dotted lines show $\rho_{xx}(T)$ at a $n=30$\%
doping level and zero incommensuration ($\Delta Q=0$)
respectively. } \label{fig:incom}
\end{figure}

\begin{figure}
\caption{The sensitivity of $\rho_{xx}(T)$ with respect to the
changes in spin-fluctuation parameters. The curves correspond to
(top to bottom)
$\omega_{sf}=20 + 0.2 T$(K), $60+0.2T$(K), $20 + 0.6 T$(K) and
$60
+ 0.6 T$(K) respectively.
All other parameters are assumed the same as in Fig.\
\ref{fig:incom}. Inset: $\sigma_{xy}(T)$  for the input
parameters
defined in the figure.
}
\label{fig:rxxwsf}
\end{figure}

\begin{figure}
\caption{(a) $\rho_{xx}(T)$ in the PG regime: the parameters
used
are: $\xi^{-1}=0.1+T/1000$(K), $\omega_{sf}\xi=50$meV, $g=1$eV
and
$n=15$\%. The solid, dashed and dotted lines correspond to
$-t^\prime/t=0.2$, 0.3 and 0.4 respectively. The effective
interaction is assumed to have incommensurate peaks at ${\bf Q}+
(\pm \Delta Q,0)$ and ${\bf Q}+
(0,\pm \Delta Q)$, where $\Delta Q=0.25\pi$. The three curves
show
a vast difference in the crossover temperature $T_0(T)$.
(b) $\rho_{xx}(T)$ in comparison with experiment (Ref.\ 36). 
The solid line corresponds to $t'=-0.3t$, with $g$
adjusted
to yield the same slope as seen experimentally. The dashed line 
corresponds to $t^\prime=-0.25t$, with the FL scattering 
included (see text).}
 \label{fig:rxx214} \end{figure}

\begin{figure}
\caption{Comparison of the calculated value of $\cot\Theta_H$
with experiment (Ref.\ 36). The solid (dashed) 
line corresponds to the result depicted by 
the solid (dashed) line in Fig.\
\ref{fig:rxx214}b.
} \label{fig:214cot}
\end{figure}

\begin{figure}
\caption{A model FS for a system with strong a-b plane
anisotropy.
The anisotropy parameter is assumed to be $a=0.5$ (see Eq.\
(\ref{eq:aniso})), the doping level is $n\approx 15$\%.
} \label{fig:fs1}
\end{figure}

\begin{figure}
\caption{The resistivities along $a$ and  $b$ crystallographic
directions as a function of temperature for the system with the
FS depicted in Fig.\ \ref{fig:fs1}, using the spin fluctuation
parameters appropriate to YBa$_2$Cu$_4$O$_8$ system. Note that there is
not only a large quantitative difference, but a qualitative
difference as well, as is seen in experiment.
The anisotropy of the distribution of the
hot regions leads to the anisotropy of the resistivities.}
\label{fig:aniso} \end{figure}
\begin{figure}

\caption{The Hall conductivity as a function of temperature for
the system depicted in Fig.\ \ref{fig:aniso}. Note that, as usual,
$\sigma_{xy}$ reflects an average contribution of the hot and cold regions
and hence is proportional to $T^{-3}$ as in the isotropic case (see
Fig.\ \ref{fig:rxx123} and its inset).
} \label{fig:hallaniso}
\end{figure}

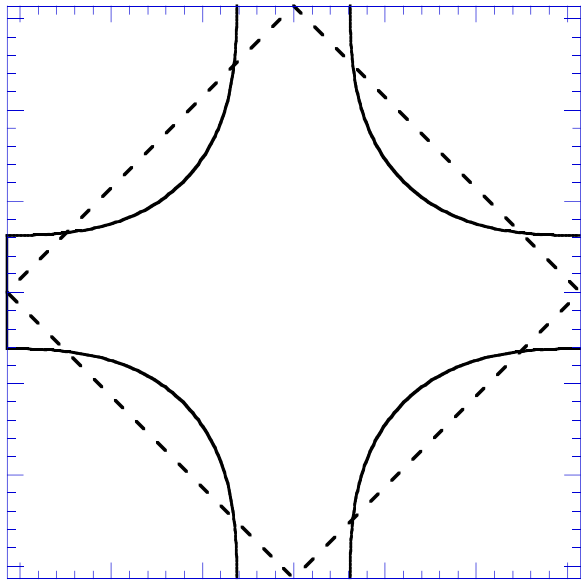
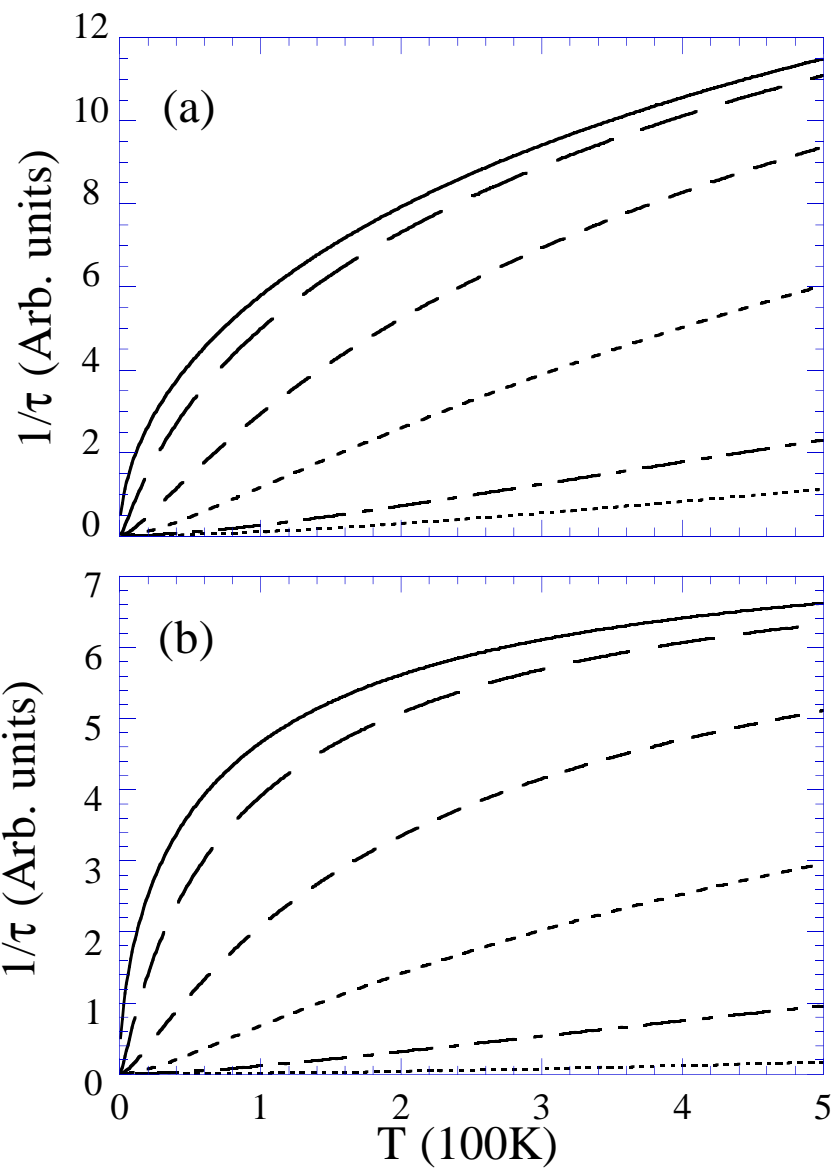
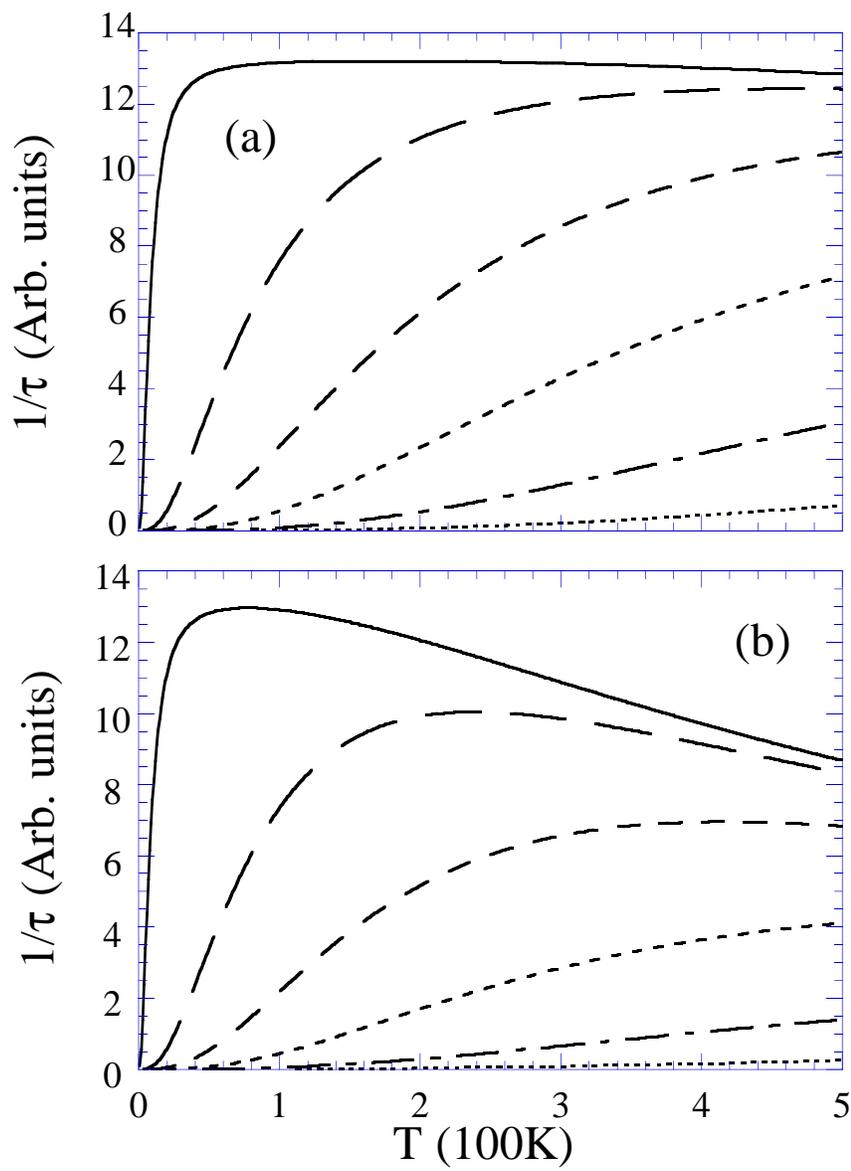
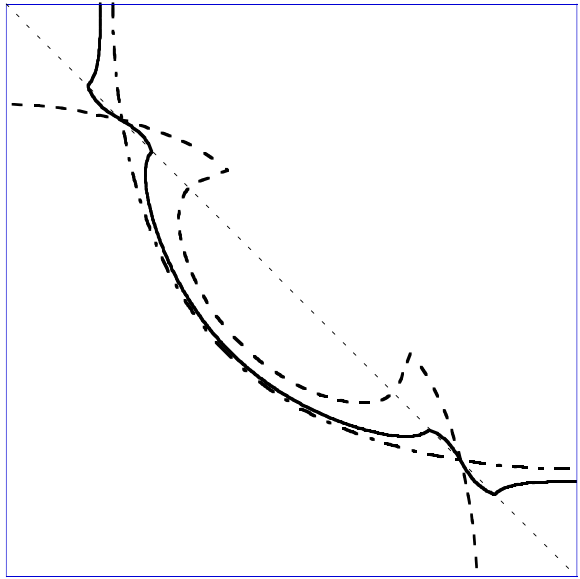
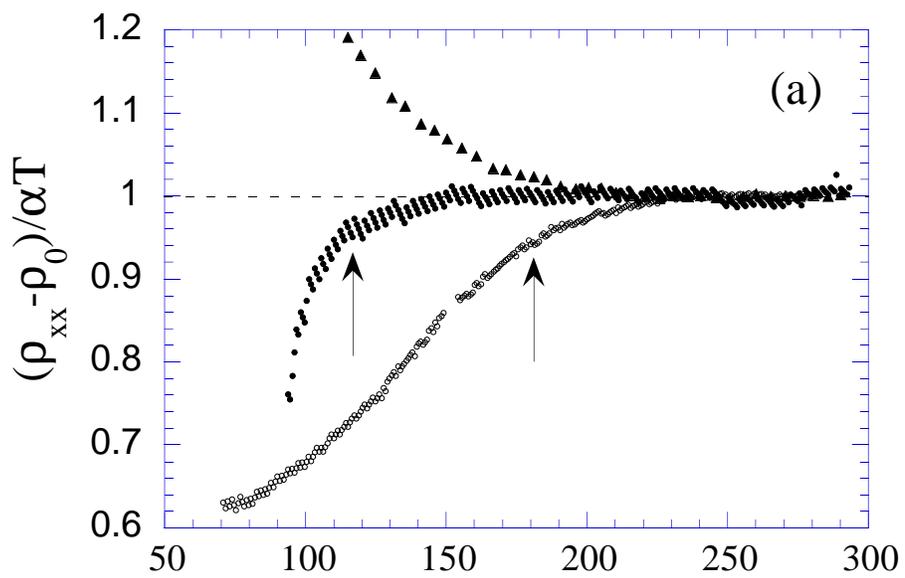
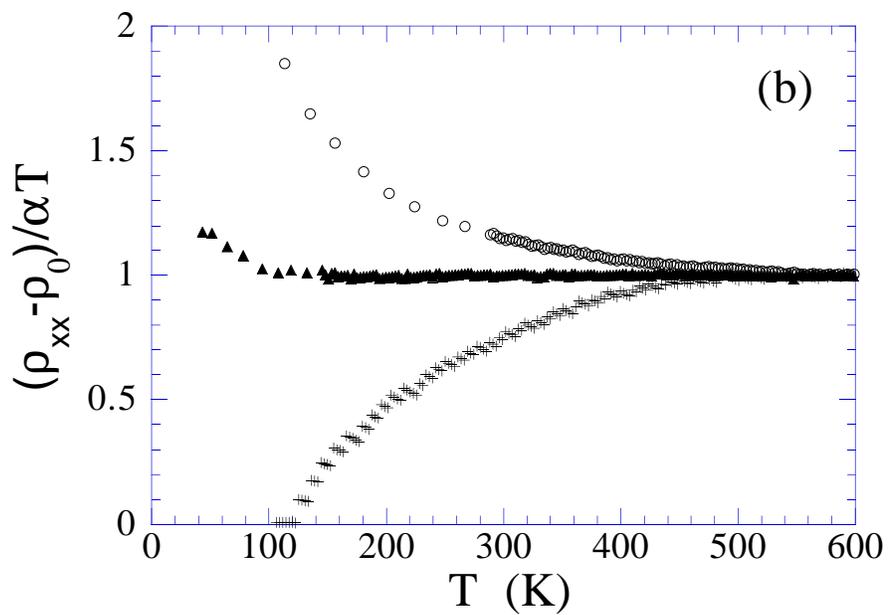
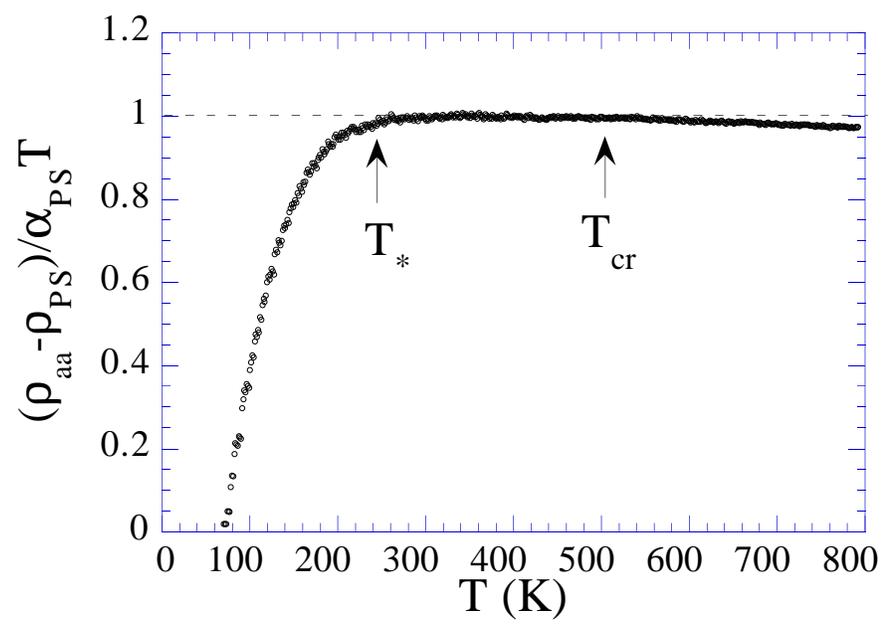
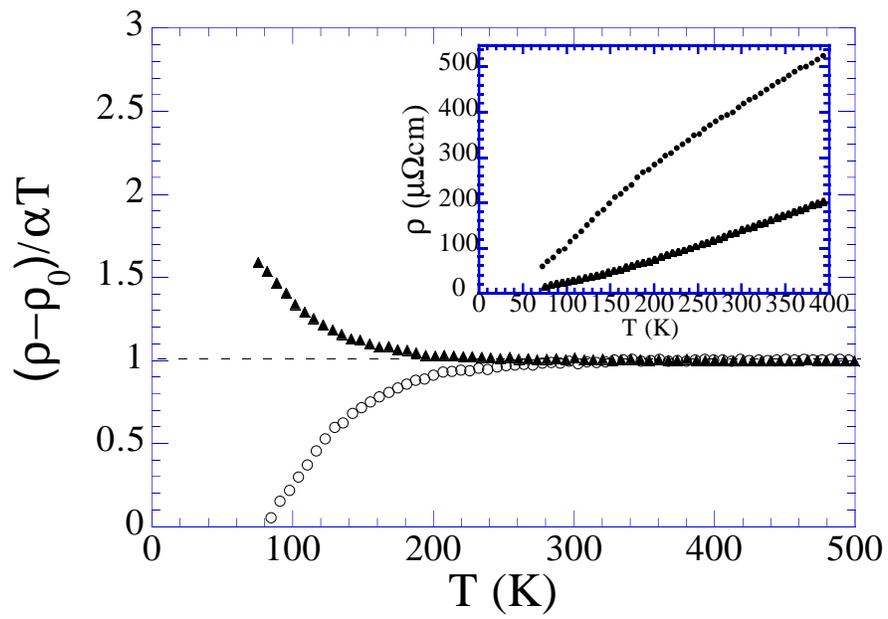
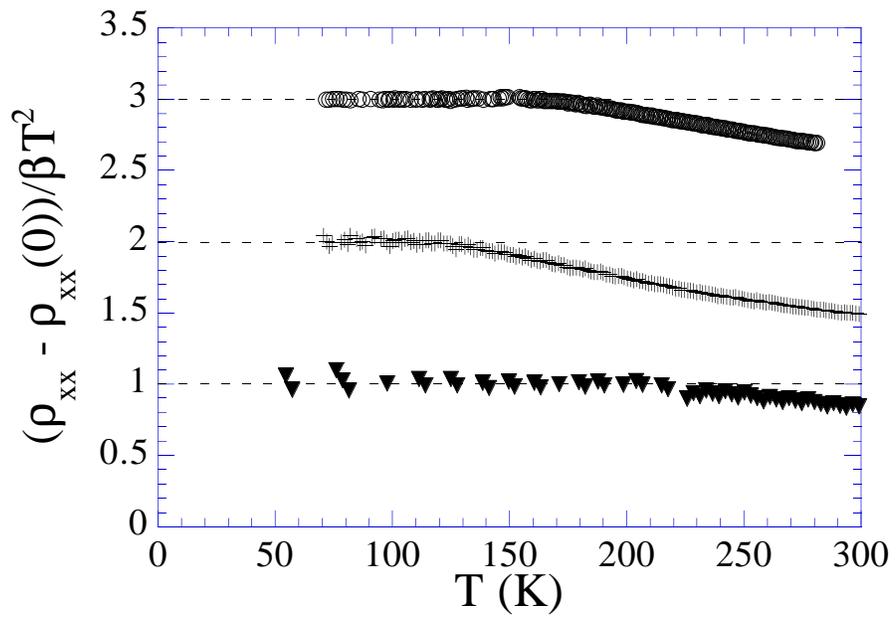
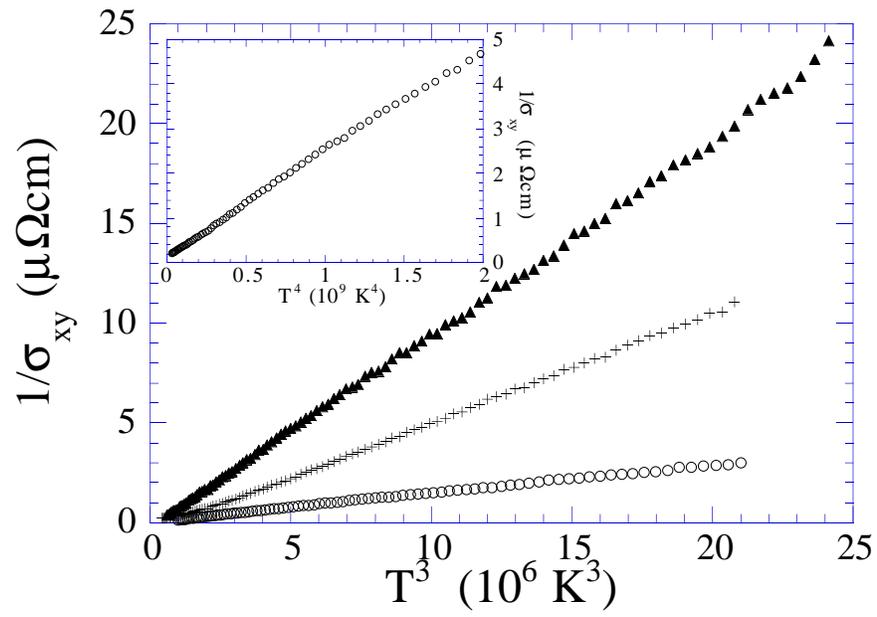
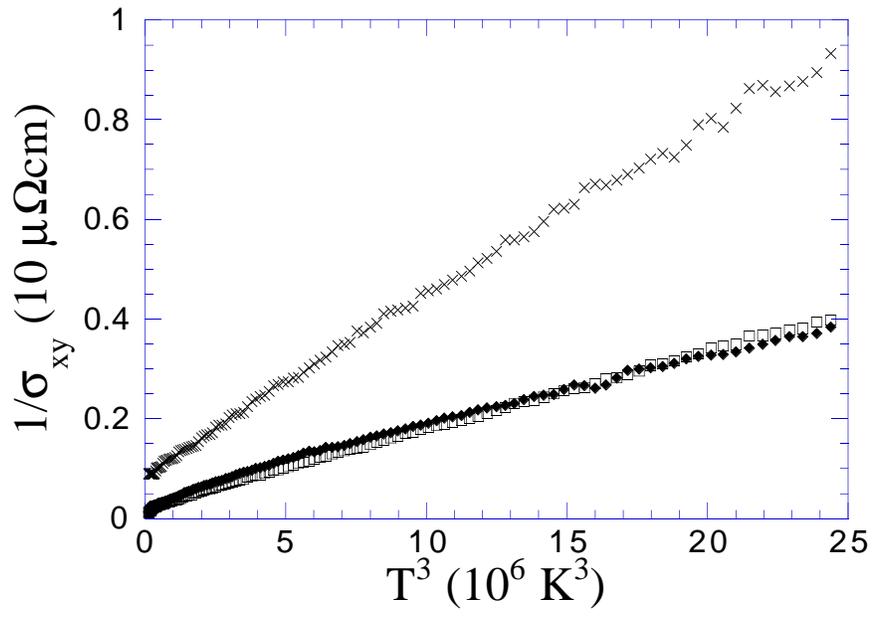
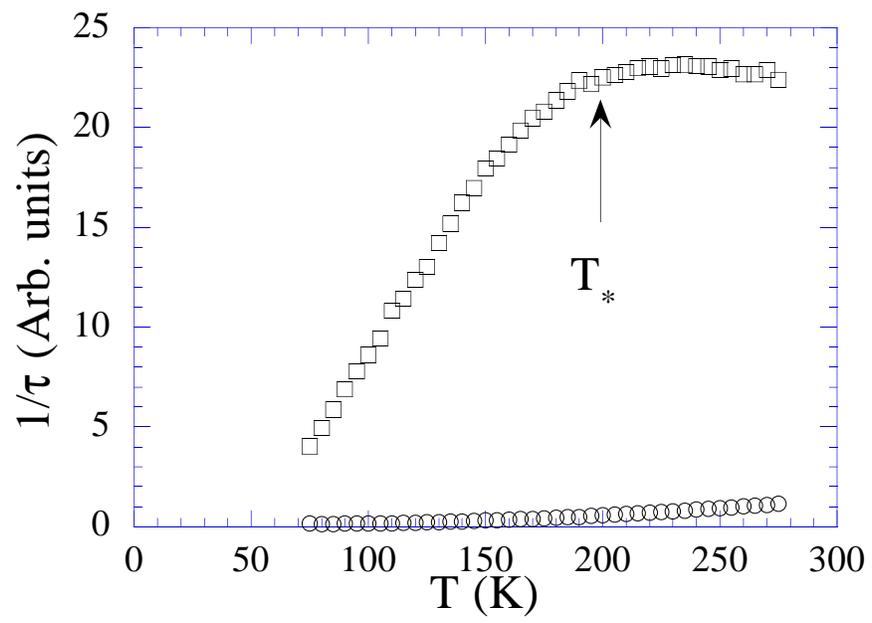
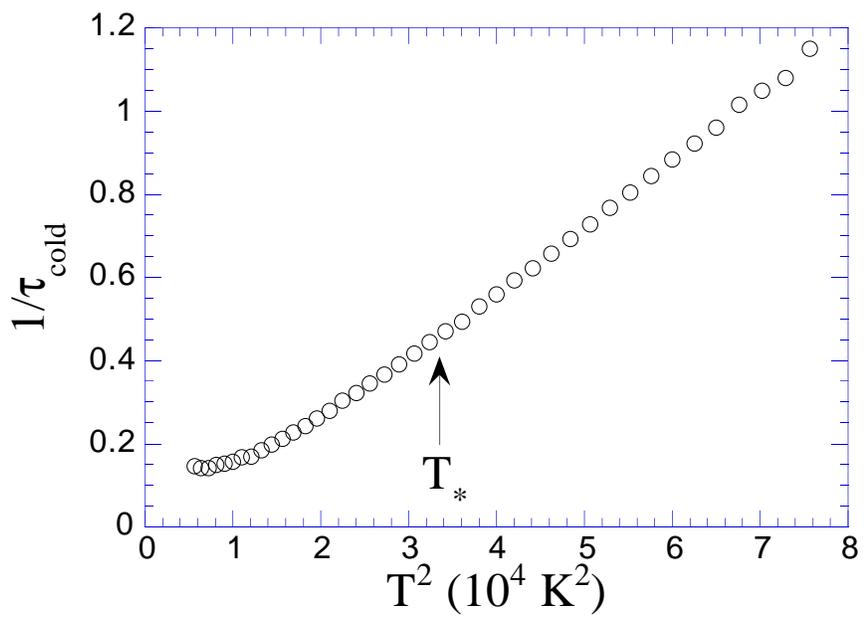
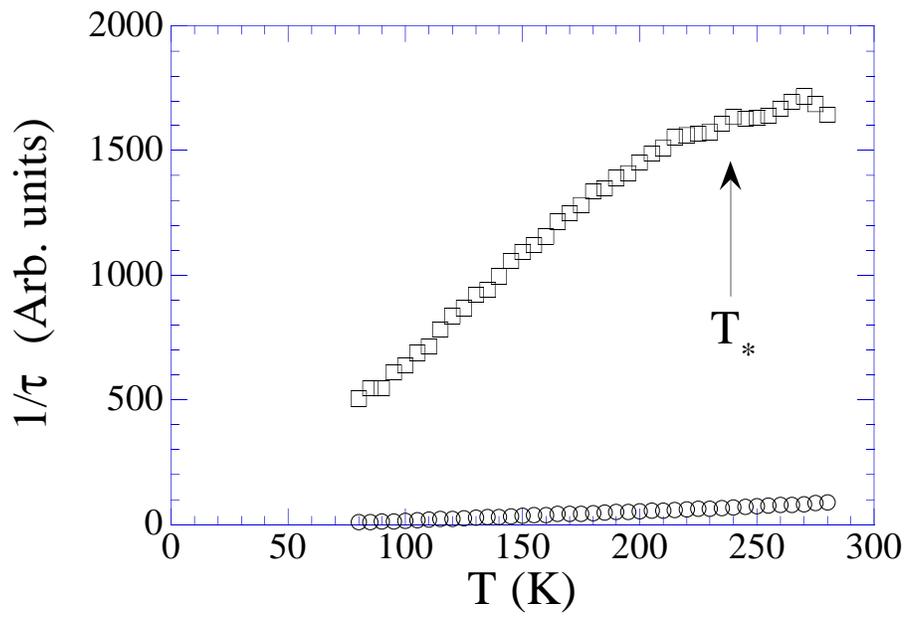
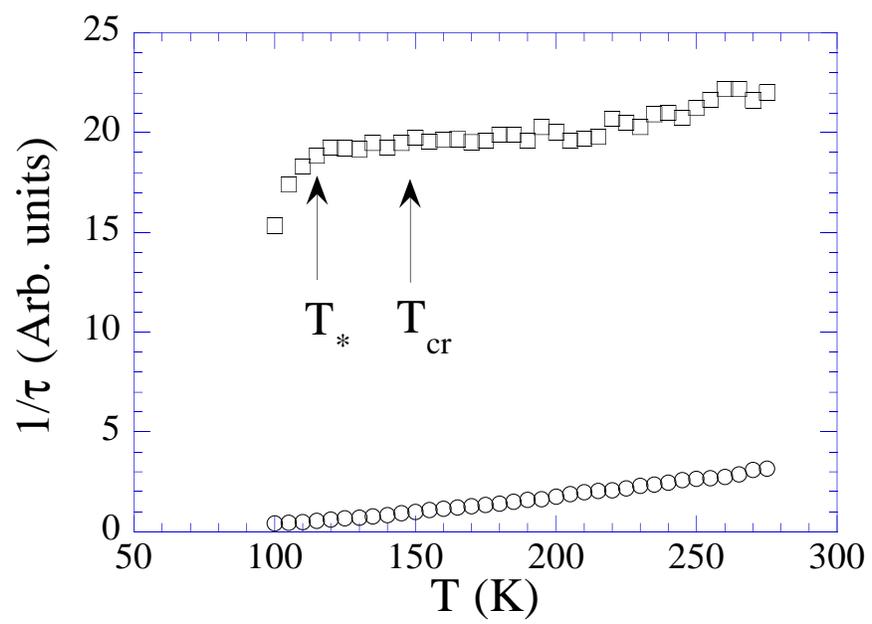
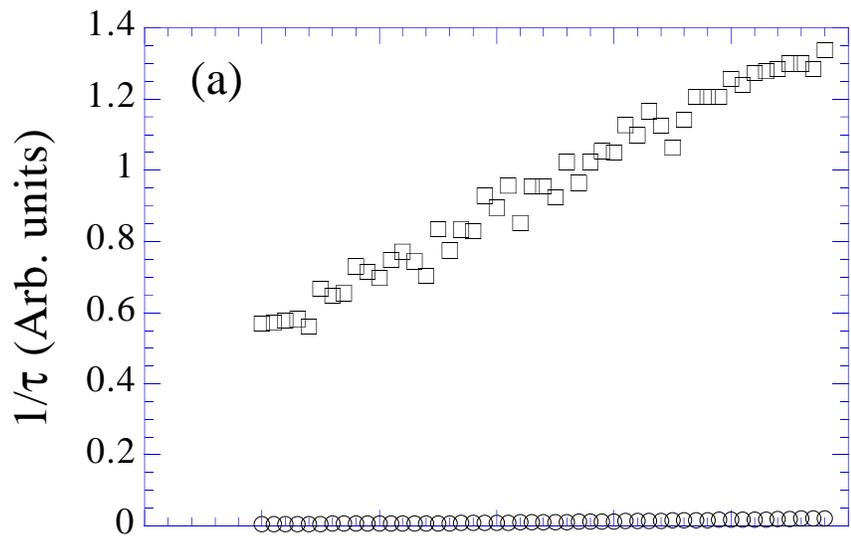
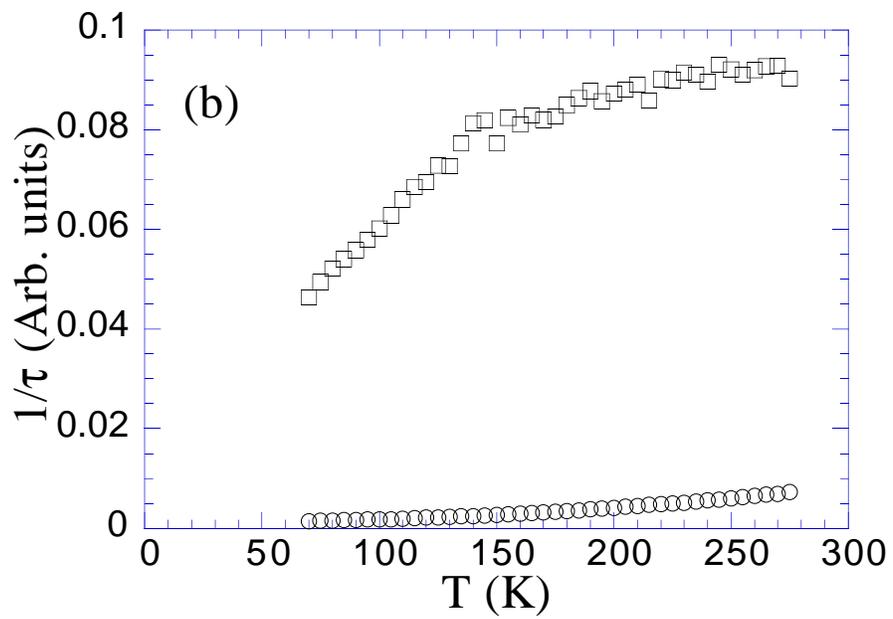
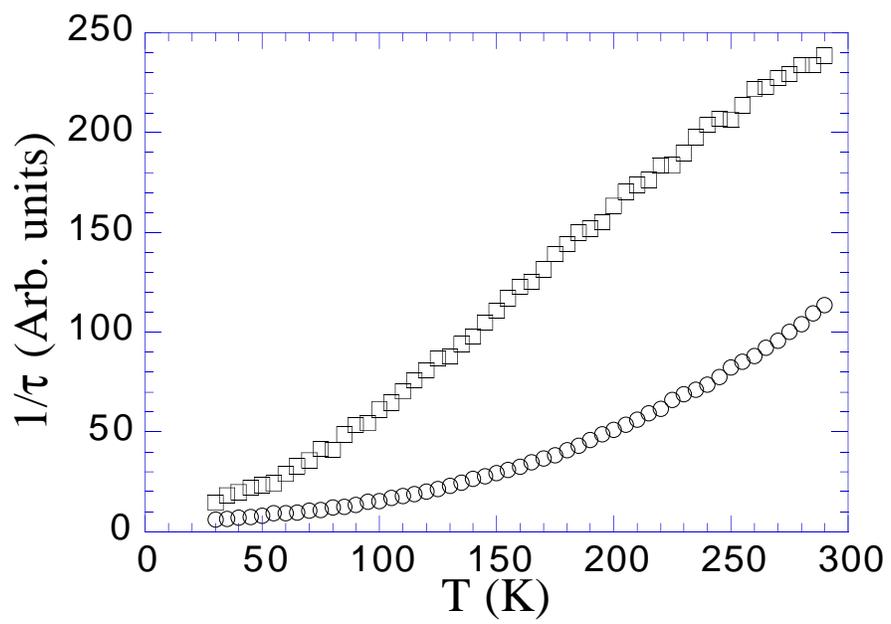
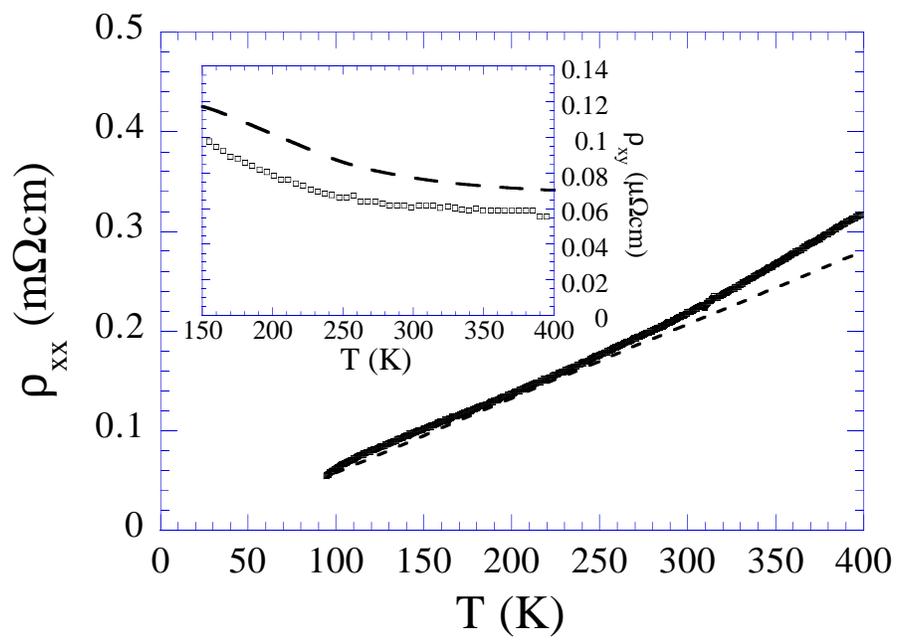
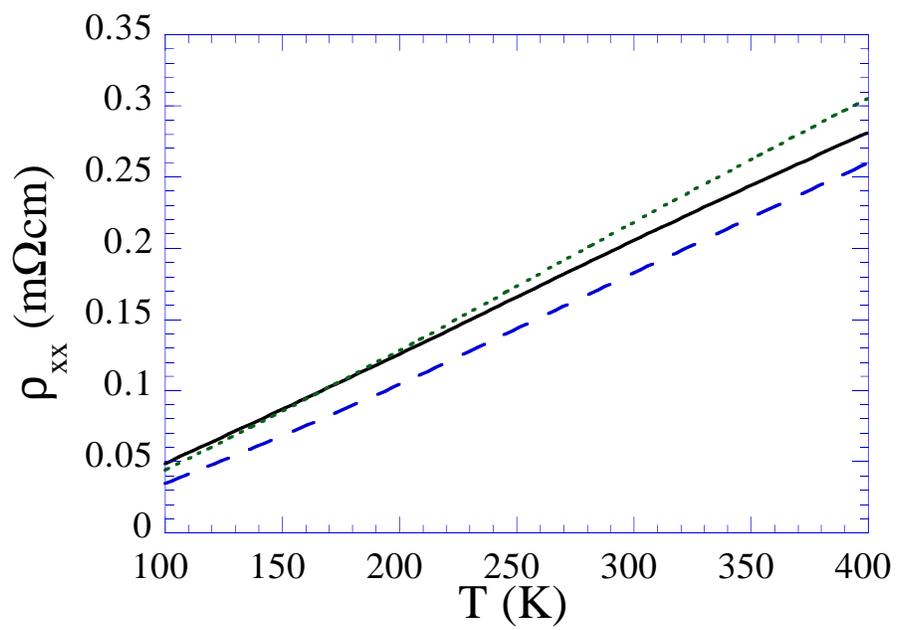
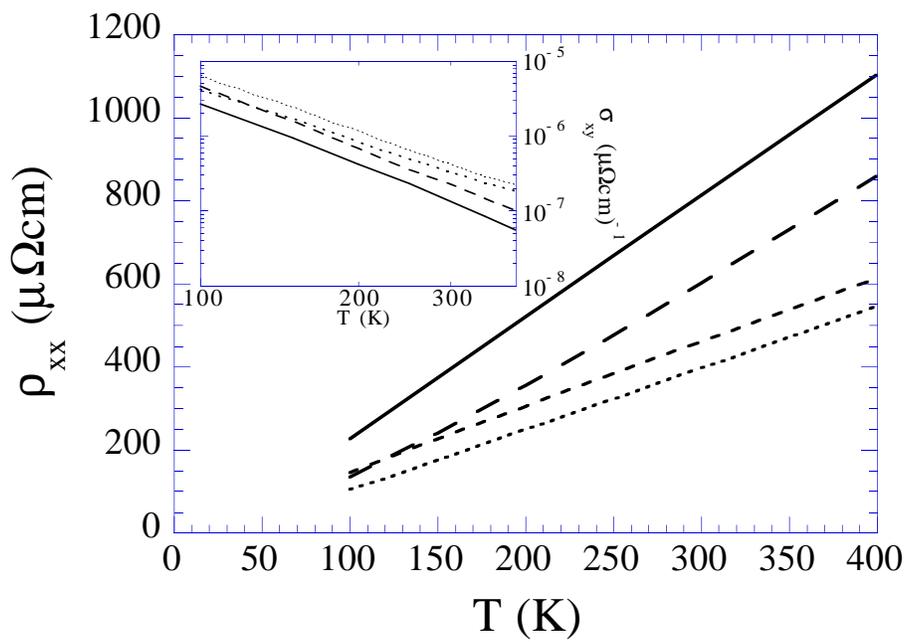
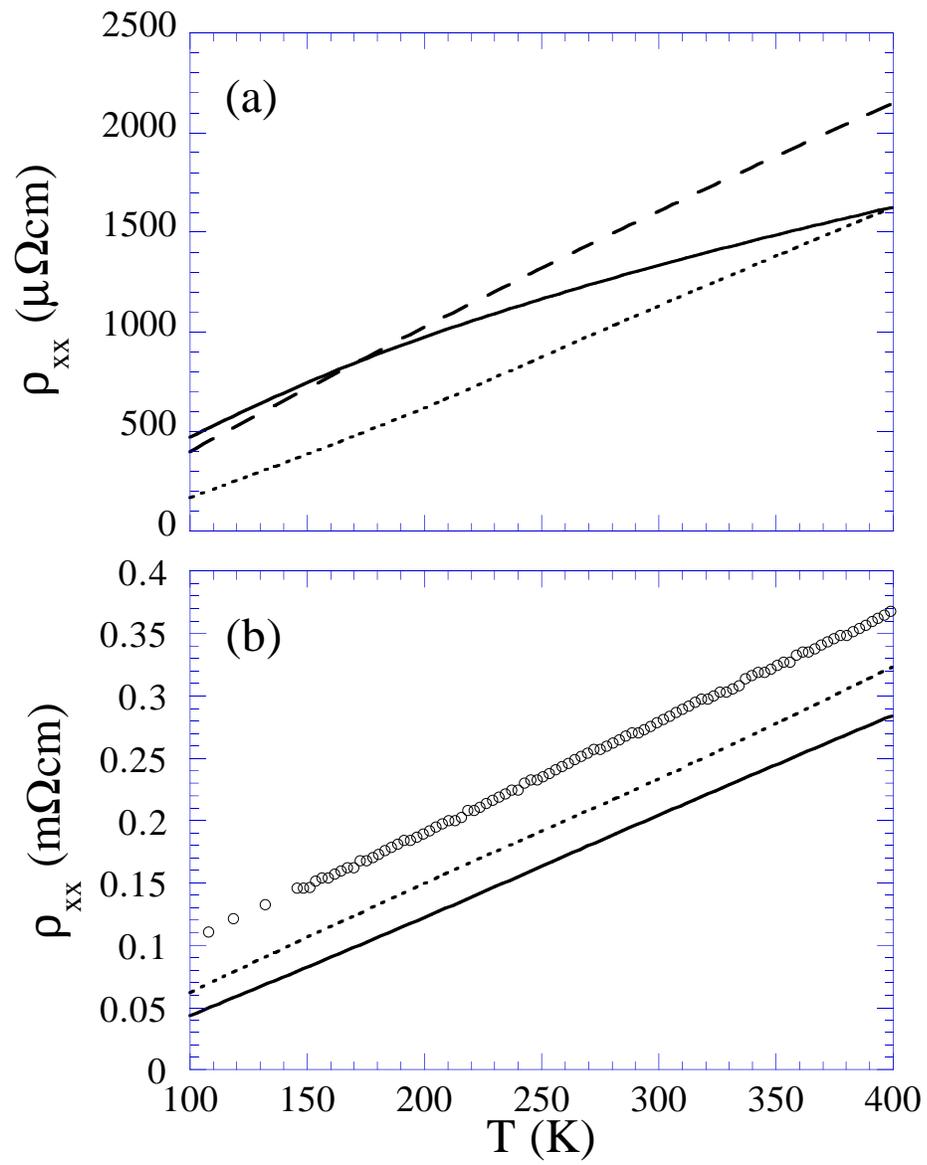
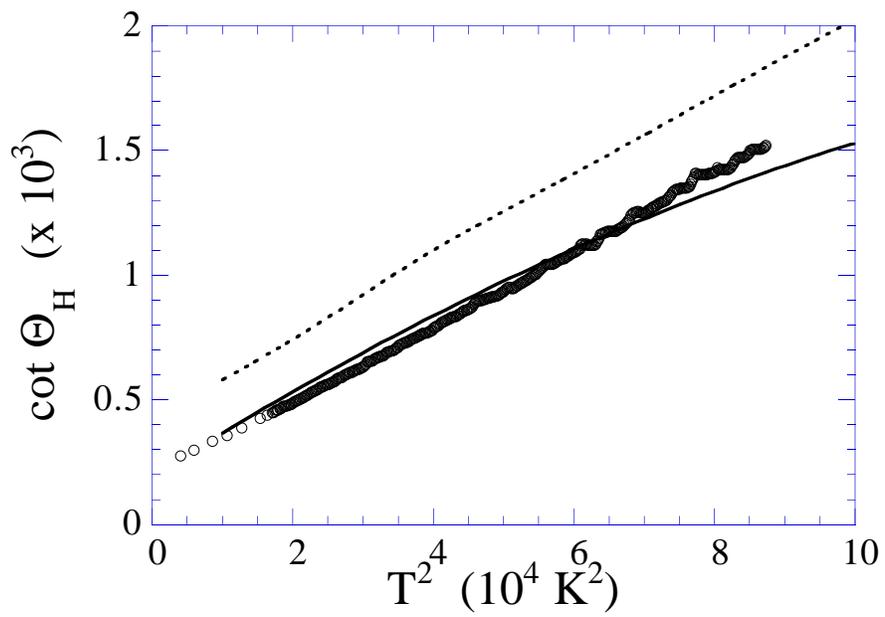
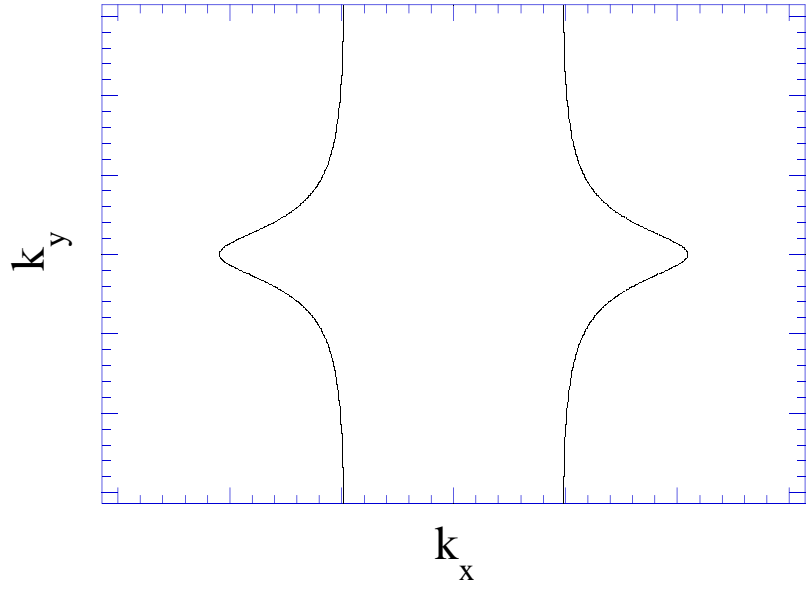
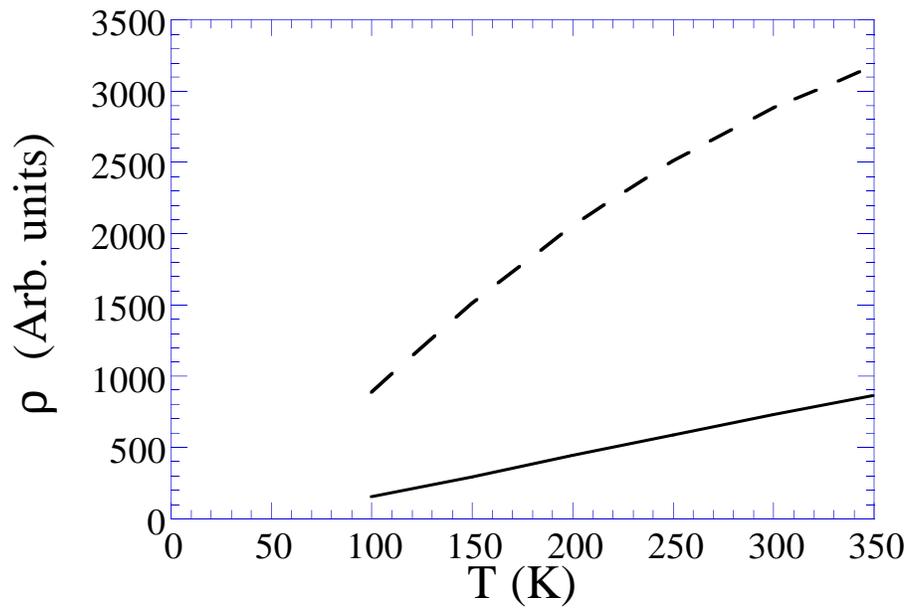
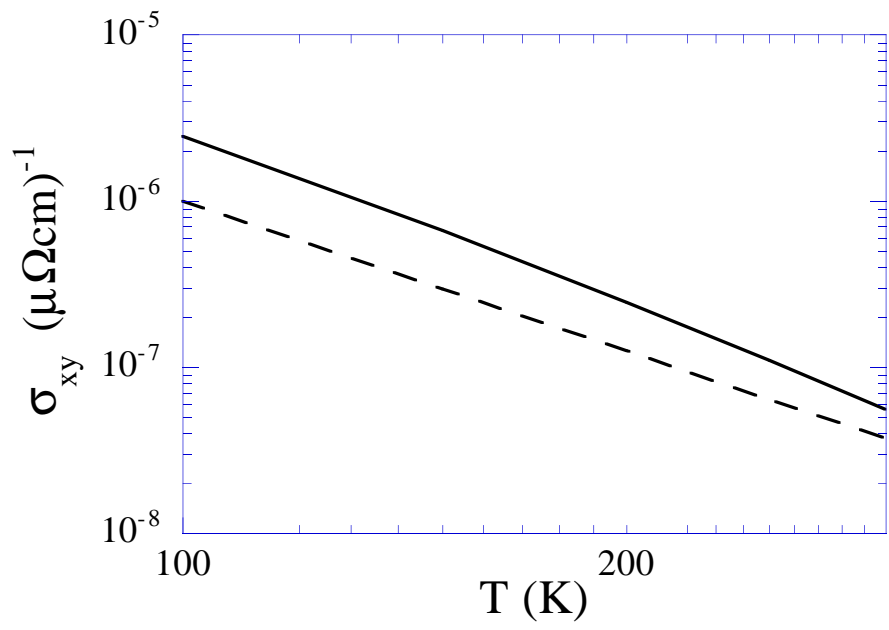
\end{document}